\documentclass[
    twocolumn,
    prb,
    showpacs,
    preprintnumbers,
    superscriptaddress,
    amsmath,
    amssymb,
    citeautoscript
    ]{revtex4-1}
    
\usepackage{graphicx}
\usepackage{epstopdf}
\usepackage{dcolumn}
\usepackage{color}
\usepackage{multirow}
\usepackage{rotating}
\usepackage{array}
\usepackage{bm}

\DeclareMathOperator{\sgn}{sgn}
\DeclareMathOperator{\re}{Re}
\DeclareMathOperator{\im}{Im}
\DeclareMathOperator{\T}{T}
\DeclareMathOperator{\Pf}{Pf}

\newcommand{\be}{\begin{equation}}
\newcommand{\ee}{\end{equation}}
\newcommand{\ba}{\begin{eqnarray}}
\newcommand{\ea}{\end{eqnarray}}
\newcommand{\bpm}{\begin{pmatrix}}
\newcommand{\epm}{\end{pmatrix}}
\newcommand{\bs }{\boldsymbol }
\newcommand{\ver}{\bs{r}}

\graphicspath{{./figs/}}

\begin{document}

\title{Majorana fermions in finite-size strips with in-plane magnetic fields}

\author{V. Kaladzhyan}
\email{vardan.kaladzhyan@cea.fr}
\affiliation{Institut de Physique Th\'eorique, CEA/Saclay,
Orme des Merisiers, 91190 Gif-sur-Yvette Cedex, France}
\affiliation{Laboratoire de Physique des Solides, CNRS, Univ. Paris-Sud, Universit\'e Paris-Saclay, 91405 Orsay Cedex, France}
\author{J. Despres}
\affiliation{Institut de Physique Th\'eorique, CEA/Saclay,
Orme des Merisiers, 91190 Gif-sur-Yvette Cedex, France}
\affiliation{Laboratoire Charles Fabry, Institut d'Optique, CNRS, Univ. Paris-Saclay, 2 avenue Augustin Fresnel, F-91127 Palaiseau Cedex, France}
\author{I. Mandal}
\affiliation{Department of Physics, University of Basel, Klingelbergstrasse 82, CH-4056 Basel, Switzerland}
\author{C. Bena}
\affiliation{Institut de Physique Th\'eorique, CEA/Saclay,
Orme des Merisiers, 91190 Gif-sur-Yvette Cedex, France}
\affiliation{Laboratoire de Physique des Solides, CNRS, Univ. Paris-Sud, Universit\'e Paris-Saclay, 91405 Orsay Cedex, France}

\date{\today}

\begin{abstract}
We study the Majorana bound states arising in finite-size strips with Rashba spin-orbit coupling in the presence of an in-plane Zeeman magnetic field. Using two different methods, first, the numerical diagonalization of the tight-binding Hamiltonian, and second, finding the singular points of the Hamiltonian (see Refs.~[\onlinecite{Mandal2015,Mandal2016a,Mandal2016b,Aguado2016}]), we obtain the topological phase diagram for these systems as a function of the chemical potential and the magnetic field, and we demonstrate the consistency of these two methods. By introducing disorder into these systems we confirm that the states with even number of Majorana pairs are not topologically protected. Finally, we show that a
calculation of the $\mathbb{Z}_2$ topological invariants recovers correctly the parity of the number of Majorana bound states pairs, and it is thus fully consistent with the phase diagrams of the disordered systems. 
\end{abstract}

\maketitle

\section{Introduction}

There has been a lot of progress recently in realizing Majorana bound states (MBS) in various low-dimensional systems both theoretically\cite{Kitaev2001,Fu2008,Fu2009,Sato2009,Lutchyn2010,Oreg2010,Potter2010,Qi2011,Martin2012,Tewari2012,Klinovaja2012,Klinovaja2013,NadjPerge2013,Sau2013,Pientka2013,Mizushima2013,Wang2014,Poyhonen2014,Seroussi2014,Wakatsuki2014,SanJose2014,Mohanta2014,BenShach2015}   and experimentally\cite{Deng2012,Mourik2012,Das2012,Lee2014,NadjPerge2014}. The key ingredients are a strong Rashba spin-orbit coupling and Zeeman magnetic fields. However, the inevitable orbital effects of the magnetic field drastically modify the topological phase diagram, eventually destroying the MBS \cite{Lim2012,Lim2013,Osca2015,Nijholt2016}. One possible way to avoid orbital effects is to use in-plane magnetic fields \cite{Kjaergaard2012,Loder2015,Albrecht2016,Li2016}. 

In Refs. [\onlinecite{Sticlet2012,Sedlmayr2015a,Sedlmayr2015b,Sedlmayr2016}] various low-dimensional systems, such as infinite ribbons and finite-size strips, have been studied. These systems respect particle-hole symmetry (PHS), but violate time-reversal symmetry (TRS) due to the presence of a magnetic field. Therefore, the corresponding topological phase diagrams have been obtained by computing the $\mathbb{Z}_2$ topological invariant (in accordance with the well-known tenfold topological classification \cite{Altland1997}), as well as numerical methods. However, only the case of magnetic fields perpendicular to the plane of the system has been considered. 

In this paper we consider the formation of MBS in infinite ribbons and finite-size strips, while focusing on in-plane magnetic fields. 
We thus show that infinite ribbons can host one or two pairs of chiral Majorana modes\cite{Sato2010,Qi2010,Wu2012,Daido2017,He2017} in the presence of an in-plane magnetic field perpendicular to the edges, and we calculate the corresponding topological phase diagram. For finite-size strips we first use a numerical diagonalization of the tight-binding Hamiltonian \cite{matq} which allows us to evaluate the Majorana polarization (MP)\cite{Sticlet2012,Sedlmayr2015b,Sedlmayr2016}. We show that one or multiple MBS pairs can form along the short edges of the system for an in-plane magnetic field parallel to the longer dimension of the finite-size strips, and we calculate the topological phase diagrams of these systems. Second we use the singular points (SP) technique introduced in Refs.~ [\onlinecite{Mandal2015,Mandal2016a,Mandal2016b,Aguado2016}], based on the momentum values where the determinant of the Hamiltonian vanishes, and we show that it yields results consistent with the numerical ones.  

We study the stability of the resulting topological states with respect to disorder\cite{Sedlmayr2015a}, and we confirm that the states with even numbers of Majorana fermions are not protected, whereas those with odd numbers are. We also perform a calculation of the $\mathbb{Z}_2$ invariant which should give one access to the parity of the number of Majorana modes. Indeed we find that this calculation predicts correctly a topologically non-trivial character for the phase-space regions shown numerically to have an odd number of MBS pairs and to survive the effects of disorder. 

The paper is organized as follows: in Sec. \ref{Model} we introduce the general model and give a concise description of the tight-binding and singular points techniques, as well as of the Majorana polarization definition. In Sec. III we present the results for 1D wires and infinite ribbons. In Sec. IV we present the phase diagrams for finite-size strips of different widths, using the numerical diagonalization and singular points methods. In Sec V we present the effects of disorder and we compare the disordered results with those obtained using a topological-invariant calculation. In Section VI we consider finite-size square systems. Finally, we conclude in Sec. \ref{Conclusions} leaving the technical details of topological-invariant calculations for the Appendix.

\section{Model and methods} \label{Model}
We introduce a model that can describe different 1D and 2D systems with an intrinsic (or proximity-induced) s-wave superconducting pairing $\Delta$, longitudinal and transversal Rashba spin-orbit couplings $\lambda_{x,y}$, and a Zeeman magnetic field $\bs{B} = (B_x, B_y, B_z)$. We write the Hamiltonian in the Nambu basis $\Psi_{\ver} = \left\{ c^\dag_{\ver \uparrow},\, c^\dag_{\ver \downarrow},\, c_{\ver \downarrow},\,  -c_{\ver \uparrow} \right\} $, where $c_{\ver \sigma}$ ($c^{\dag}_{\ver \sigma}$) annihilates (creates) a particle of spin $\sigma$ at site $ \ver = (i,j)$ in a square lattice:
\begin{eqnarray}
H = &\sum\limits_{\ver}\left[ \Psi^\dag_{\ver} \left(-\mu \tau_z + \Delta \tau_x + \bs{B \cdot \sigma}\right) \Psi_{\ver} \right. + \\ \nonumber
&+ \Psi^\dag_{\ver} \left(-t_x -i \lambda_x \sigma_y \right) \tau_z \Psi_{\ver + \bs{x}} + \mathrm{H.c.} + \\ \nonumber
&+ \left. \Psi^\dag_{\ver} \left(-t_y +i \lambda_y \sigma_x \right) \tau_z  \Psi_{\ver+\bs{y}}+ \mathrm{H.c.} \right],
\label{TBHamil}
\end{eqnarray}
where $t$ is the hopping amplitude, $\mu$ denotes the chemical potential, $\bs{x}$,$\bs{y}$ are unit vectors for the $x$ and $y$ directions correspondingly, and the lattice spacing is set to unity. 



\subsection{Numerical tight-binding techniques and the Majorana polarization}

The eigenstates of the tight-binding Hamiltonian described above can be obtained using a numerical diagonalization (here performed using the MatQ code\cite{matq}). In the Nambu basis, an eigenstate $j$ of the tight binding Hamiltonian can be written as $\psi^{j\T}_{\ver} =  \left\{ u^j_{\ver \uparrow},\, u^j_{\ver \downarrow},\, v^j_{\ver \downarrow},\,  -v^j_{\ver \uparrow} \right\}$, where $u$ and $v$ denote the electron and hole components respectively. The vector of the local Majorana polarization \cite{Sticlet2012,Sedlmayr2015b,Sedlmayr2016} on each site $\ver = (x,y)$, for the eigenstate $j$ is given by:
\begin{equation}
P^j({\ver}) \equiv \begin{pmatrix} P^j_x({\ver}) \\ P^j_y({\ver}) \end{pmatrix} \equiv \begin{pmatrix} -2 \re \left[u^j_{\ver \uparrow} v^j_{\ver \uparrow} + u^j_{\ver \downarrow} v^j_{\ver \downarrow} \right] \\ -2 \im \left[u^j_{\ver \uparrow} v^j_{\ver \uparrow} + u^j_{\ver \downarrow} v^j_{\ver \downarrow} \right] \end{pmatrix} 
\label{MPvector}
\end{equation}
This quantity allows to discriminate locally pure electron (hole) states from Majorana-like states. It is easy to see that for pure electron ($v_{\ver \uparrow}=v_{\ver \downarrow}=0$) and pure hole states ($u_{\ver \uparrow}=u_{\ver \downarrow}=0$) the local Majorana polarization equals to zero. For our purposes it is more practical to use the integral of the MP vector over a spatial region $\mathcal{R}$ defined as:
 \begin{equation}
C_j = \left| \sum\limits_{\ver \in \mathcal{R}} \left[ P^j_x({\ver}) + i P^j_y({\ver})\right] \right|^2
\label{MPint}
\end{equation}
Note that in Eqs. (\ref{MPvector},\ref{MPint}) we assume that the wave function is normalized. 

To obtain the topological phase diagram we first find the lowest energy states of the given system. If these states have energies close to zero they may be MBS. We divide our system into two halves (along the shorter length), and we compute the integral of the MP vector in each of these halves defined by $\ver \in \mathcal{R}$ for each 'zero'-energy state. The states that have $C=1$ are MBS, and those with $C=0$  are regular electron or hole states. Note that we may have only a pair of state with $C=1$, or multiple degenerate zero-energy MBS states with $C=1$. In the results that we present here we sum the MP over all the lowest energy states. Note however that the states with even $C$ (even number of MBS pairs) are not topologically protected, and thus any small disorder introduced into the system destroys such Majorana states.

\subsection{Singular points of the Hamiltonian}

To find a MBS in a system described by a PHS Hamiltonian $H(k)$, we seek for localized zero-energy solutions of the Schr\"odinger equation: $H(k) \Phi = 0$. In the most general case these solutions are of the form $e^{i k r}$ that can be rewritten as $e^{i k_\parallel r_\parallel} \cdot e^{-z r_\perp}$, where $k_\parallel$ denotes the ``good" quantum number and $z$ is defined below. We analytically continue $k$ to the complex plane and we consider the solutions of the following equation
\begin{equation}
\det H(k) = 0,
\label{EPequation}
\end{equation}
defining the so-called singular points $k_i \equiv k_\parallel + i z$ in the complex plane at which the determinant of the Hamiltonian vanishes. By definition $z$ is given by the imaginary part of the singular point $k_i$. The practical use of these complex momentum values is the following: by continuously changing the parameters of our Hamiltonian, we continuously change the corresponding $k_i$'s. If we are in a topological phase, $z$ must be positive (in other words the solution is localized). As soon as $z$ crosses zero and becomes negative, the solution becomes delocalized, and therefore we enter a non-topological phase. Further details can be found in Refs.~[\onlinecite{Mandal2015,Mandal2016a,Mandal2016b}].

We propose the following way of constructing a phase diagram. The parameter space is given by the chemical potential $\mu$ and the magnetic field $B = |\bs{B}|$. Firstly, we find all the $k_i$'s as a function of the parameters in the Hamiltonian, such that $k_i = k_i(\mu, B), i \in \overline{1,2N}$, where $2N$ is the total number of $k_i$ solutions. We then sort them  at each point in the parameter space with respect to their imaginary parts as follows: $\im k_i < \im k_{i+1}, i \in \overline{1,2N-1}$. Sorting is one way to construct continuous functions $\im k_i(\mu, B)$ in the parameter space. Although it is possible to deal with  discontinuities analytically\cite{Mandal2016b}, in numerical simulations the continuity of $ \im k_{i}$'s becomes crucial since it is hard to discriminate the zeros of $\im k_i(\mu, B)$ from its discontinuities. Subsequently we look for the functions $k_i(\mu, B)$ whose imaginary part crosses zero. This can be done by plotting their imaginary part as a function of the system parameters. Since Eq.~(\ref{EPequation}) yields pairs of solutions with opposite imaginary parts, and since these pairs have been sorted according to their imaginary parts, it is therefore sufficient to plot the imaginary part of either the smallest positive root ($i=N+1$) or of the largest negative root ($i=N$). The set of points $(\mu_0, B_0)$ where $\im k_{N}(\mu_0, B_0) = 0$ (or equivalently $\im k_{N+1}(\mu_0, B_0) = 0$) yield thus the phase transition lines between the topological and non-topological regions in the phase diagram.


Note that this technique can in principle used also to count the number of MBS present in a given phase. The corresponding counting formula is extremely simplified when ''Exceptional points'' are present\cite{Mandal2015,Mandal2016a} for a system with unbroken chiral symmetry.
In such a case, the Hamiltonian can be brought to a block off-diagonal form,
however, when the chiral symmetry is broken, this block-off diagonal form cannot be achieved in any basis and it becomes cumbersome to isolate two sets of $N$ continuous  solutions $k_i$ (with opposite imaginary parts) and plug them into the counting formula to get the total number of MBS. Thus we will use the singular points method only to find the phase-transition lines. In order to find the number of MBS pairs in a given phase we will rely on the numerical tight-binding calculations.

\section{1D wires and infinite ribbons}
\subsection{1D wire}
\begin{figure}
	\includegraphics*[width = 0.7\columnwidth]{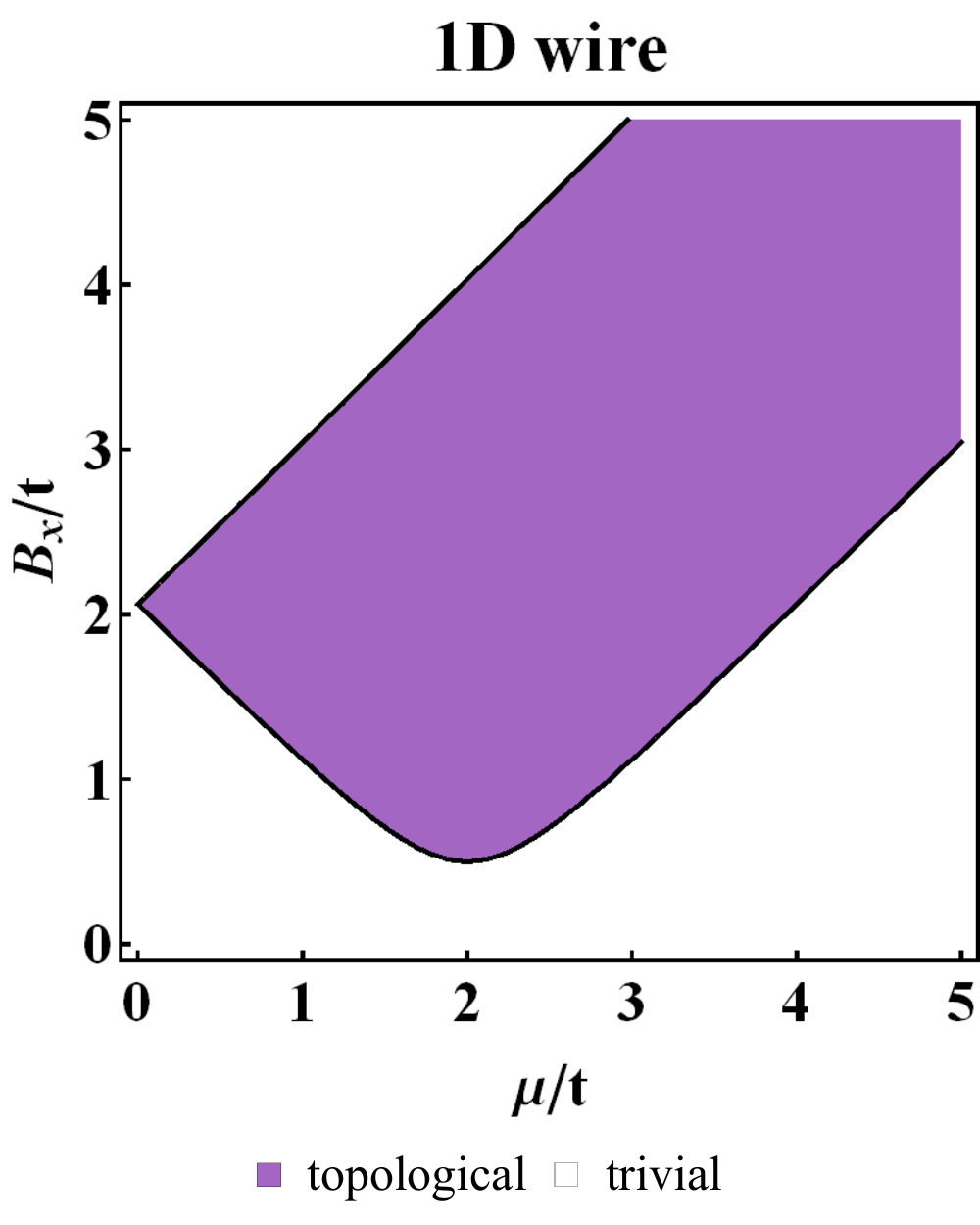}
	\caption{(color online) The phase diagram of a 1D superconducting nanowire obtained with topological invariant calculation as a function of the chemical potential $\mu$ and the magnetic field along the wire $B=B_x$ (the phase diagram remains the same in the case of a magnetic field perpendicular to the wire $B=B_z$). We set $\Delta=0.2t, \lambda_x=0.5t$.}
	\label{1wire-PhD}
\end{figure}

We start by describing the well-known phase diagram of a 1D SC wire which we take to be lying along the $x$-axis ($N_y=1$ and $N_x \gg1 $). 
In the presence of a magnetic field the time-reversal symmetry (TRS) is broken, and only the particle-hole symmetry (PHS) holds, therefore the system is in the topological class D described by a $\mathbb{Z}_2$ invariant \cite{Altland1997}. If the applied magnetic field is perpendicular to the spin-orbit direction, i.e. either $\bs{B} = (0, 0, B_z)$ or $\bs{B} = (B_x, 0, 0)$, the SC wire enters a gapful topological phase as soon as $B_x$ or $B_z$ become larger than  $\sqrt{(\mu-2t_x)^2 + \Delta^2}$. The corresponding phase diagram is shown in Fig. \ref{1wire-PhD}. Further details of the $\mathbb{Z}_2$ invariant calculation can be found in the first subsection of Appendix A.

\subsection{Infinite ribbon}
\begin{figure}[h!]
	\centering
	\includegraphics*[width = 0.8\columnwidth]{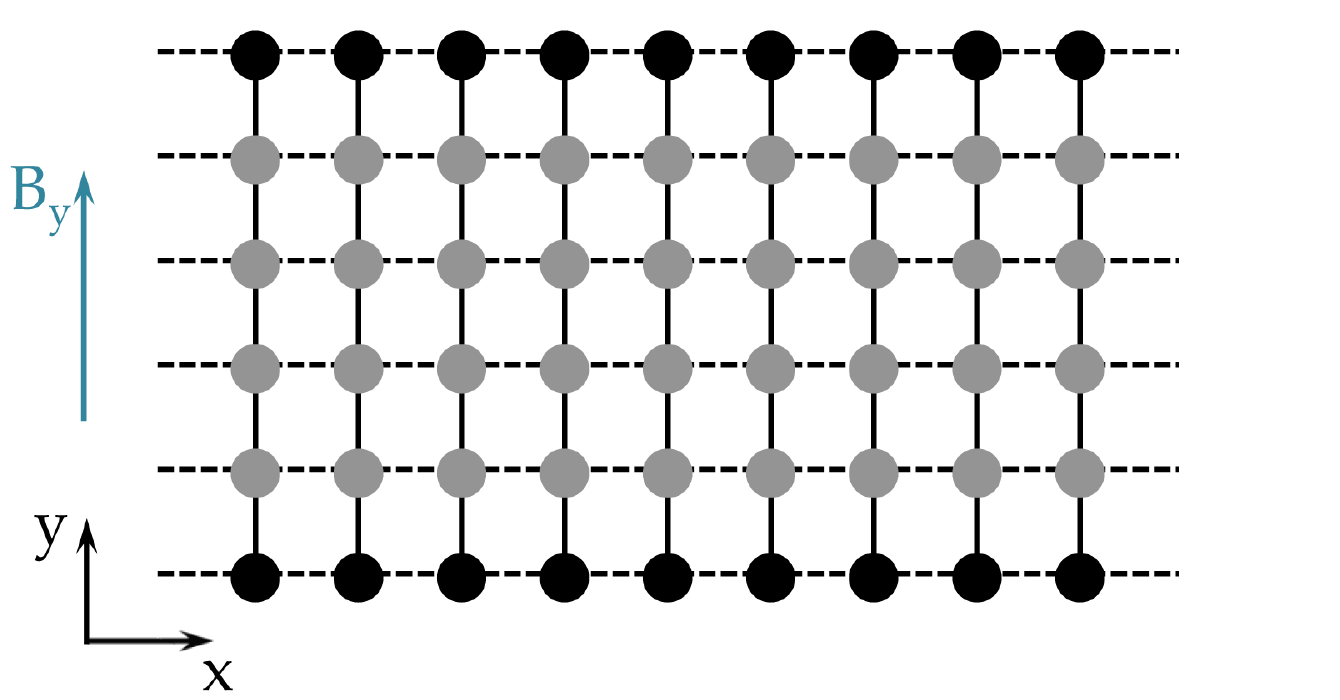}
	\caption{(color online) A sketch of an infinite ribbon along $x$-axis with a magnetic field $B=B_y$ perpendicular to its edges. The black sites denote the edges of the ribbon where the chiral Majorana modes are localized.}
	\label{2Dribbonsketch}
\end{figure}

\begin{figure}[h!]
	\centering
	\includegraphics*[width = 0.6\columnwidth]{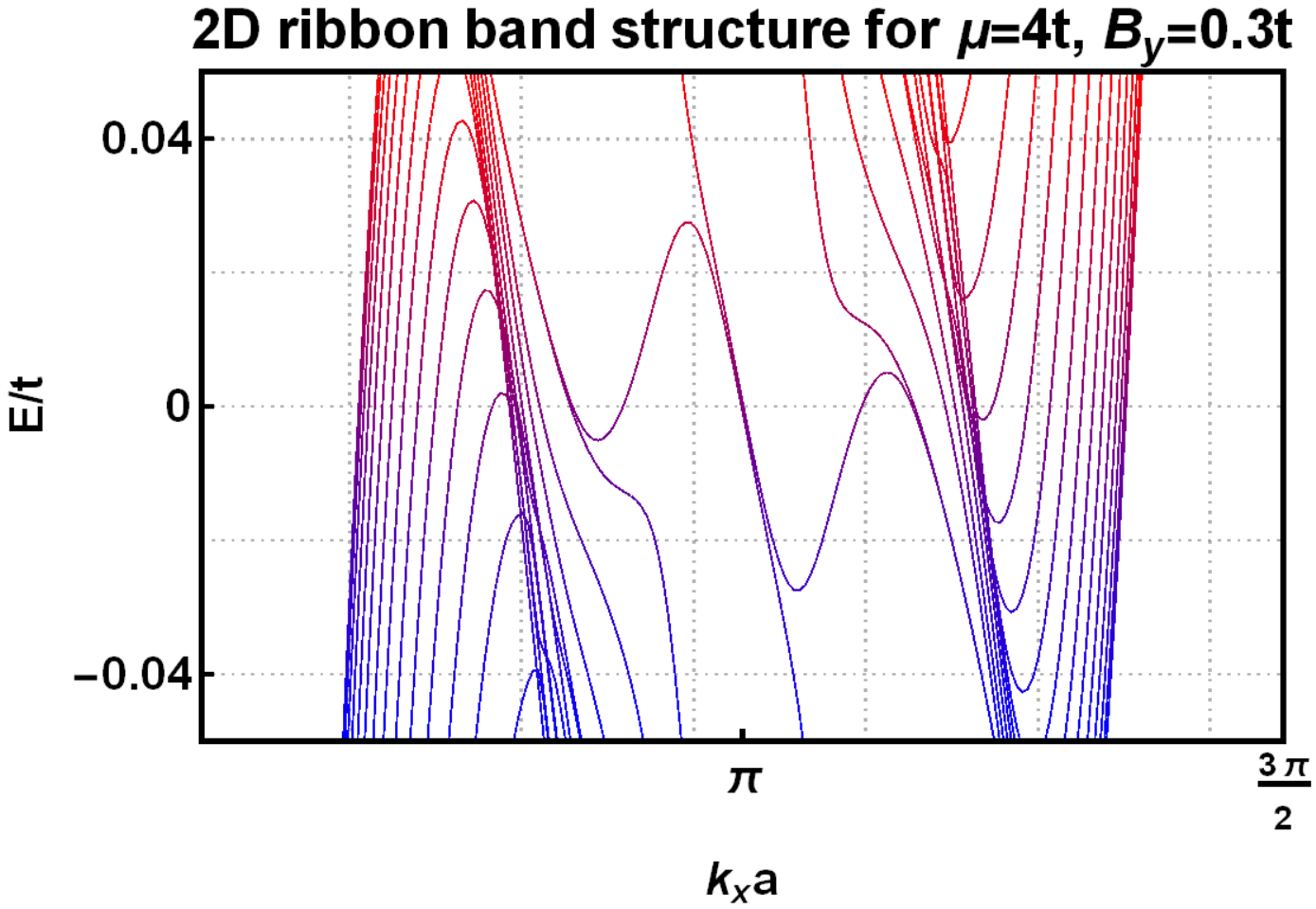}\\
	\vspace{.2in}
	\includegraphics*[width = 0.6\columnwidth]{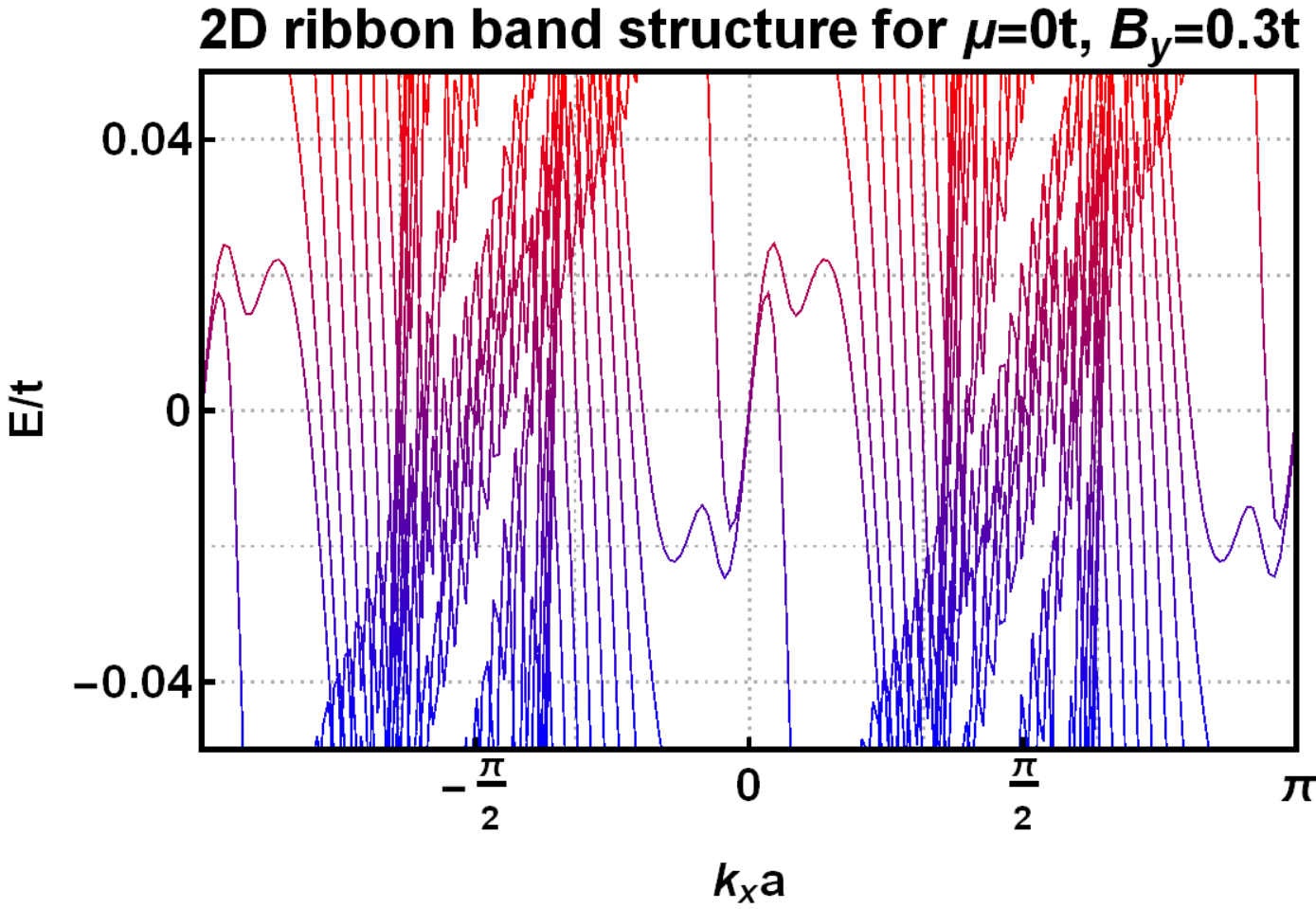}\\
	\vspace{.2in}
	\hspace{.2in}
	\includegraphics*[width = 0.7\columnwidth]{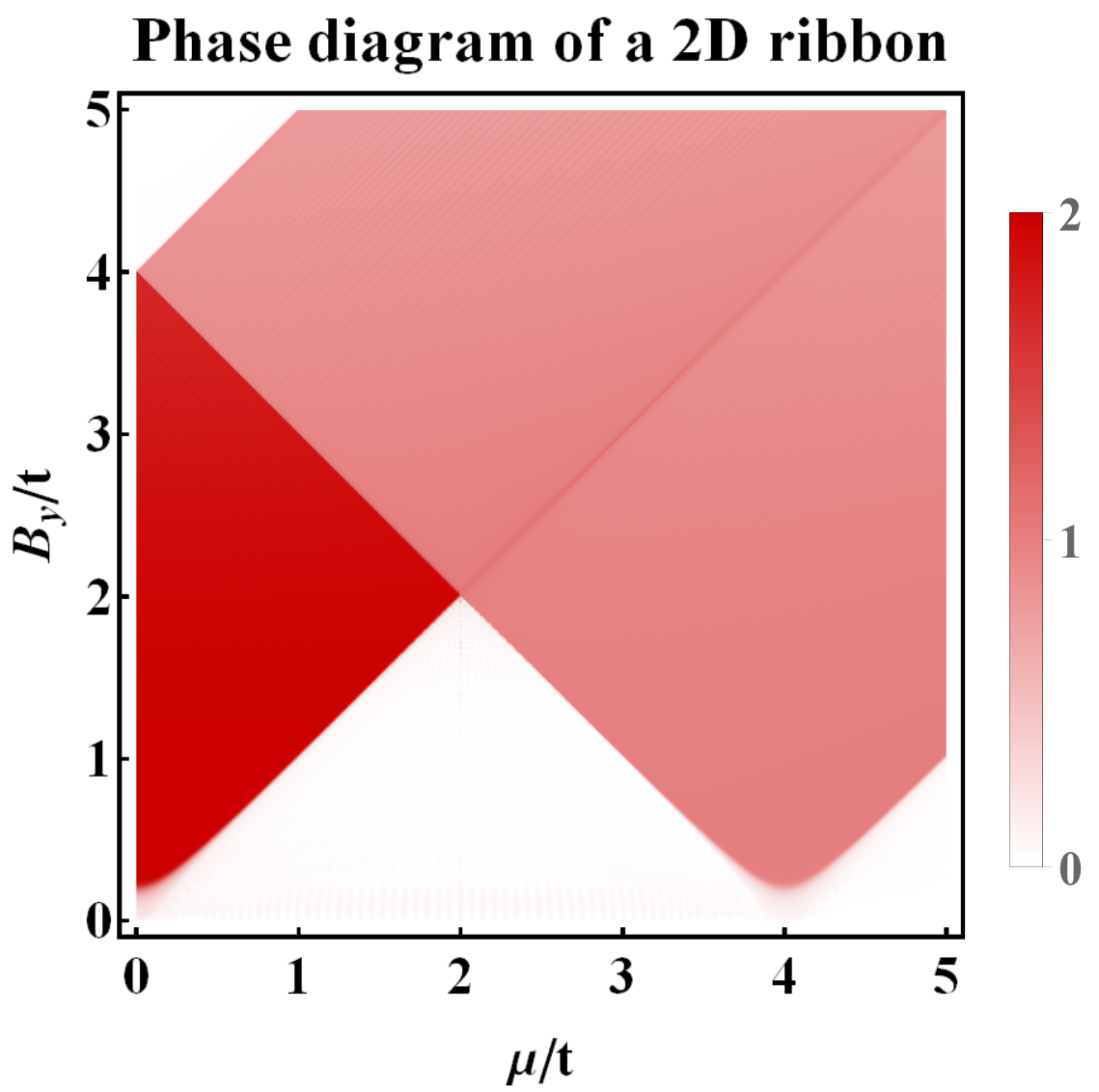}
	\caption{(color online) Band structure of an infinite ribbon with a magnetic field $B=B_y$ perpendicular to the edges for $\mu=4t, B_y=0.3t$ (upper panel) and $\mu=0t, B_y=0.3t$ (middle panel). Note that the system may host either one or two pairs of chiral Majorana modes. The corresponding topological phase diagram (lower panel) depicts the number of Majorana modes (as evaluated from the total MP) as a function of $\mu$ and $B$. In all the examples we set $\Delta=0.2t, \lambda_x=\lambda_y=0.5t$.}
	\label{2Dribbon}
\end{figure}

In this subsection we study an infinite ribbon with a finite but large number of sites in the $y$-direction ($N_y \gg 1$), and infinite in the $x$-direction $N_x\rightarrow \infty$ (see Fig.~\ref{2Dribbonsketch}). We set $\lambda_x = \lambda_y = \lambda$ and $t_x = t_y = 1$. We are interested in the formation of zero-energy edge states parallel to the $x$ axis (see the black circles in Fig.~\ref{2Dribbonsketch}). We consider open boundary conditions (also referred to as "zero boundary conditions") at the edges of the ribbon (in the $y$ direction), and that the ribbon is infinite (or equivalently periodic boundary conditions) in the $x$ direction. In Fig.~\ref{2Dribbon} we plot the band structure of this system for an in-plane magnetic field $B_y$ parallel to the $y$-axis (perpendicular to the ribbon edges), as well as the topological phase diagram of such a ribbon obtained using the tight-binding numerical diagonalization and the evaluation of the MP as described in section II. 
First of all, we note that the spectrum is PHS even though the band structure is not. Second, as we can see from the band structure, the system may become gapless, i.e. there are region in the momentum space in which the gap in the spectrum is closing. However, despite the fact that there is no overall gap, chiral MBS\cite{He2017} do form, and they correspond to values of momenta for which the bulk spectrum remains gapped (e.g. $k_x a=0$ and $k_x a=\pi$). Such states are dispersive and propagate along the edges of the ribbon. 
We should note that similar situations in which the closing the gap can occur for certain regions in the parameter space have been previously studied, and it has been shown that, despite the absence of an overall gap, the system can still be topological, and support MBS\cite{Teo2010,Matsuura2013,Deng2014,Baum2015a,Baum2015b}. 
The number of MBS pairs varies from 0 to 2, depending on the parameters of the system (see Fig.~\ref{2Dribbon}). However, the case of two Majorana fermions propagating at the same boundary is not stable, and in the absence of protection by TRS, for example in the presence of small disorder, such states would combine to form a conventional fermionic state. Thus the system is topologically non-trivial only when the number of MBS pairs is equal to 1. It is also worth mentioning that the number of sites in the $y$ direction must be large enough so that the overlap of the wave functions of the two Majorana states localized on the two opposite edges of the ribbon is exponentially small, and that these states cannot hybridize and acquire a finite energy.

Note that if the magnetic field is applied along the $x$-axis the system is also gapless, however, in this case no Majorana modes form, for any region in the parameter space, and the system is fully trivial.

\section{Finite-size strips}

\begin{figure}[h!]
	\centering
	\includegraphics*[width = 0.9\columnwidth]{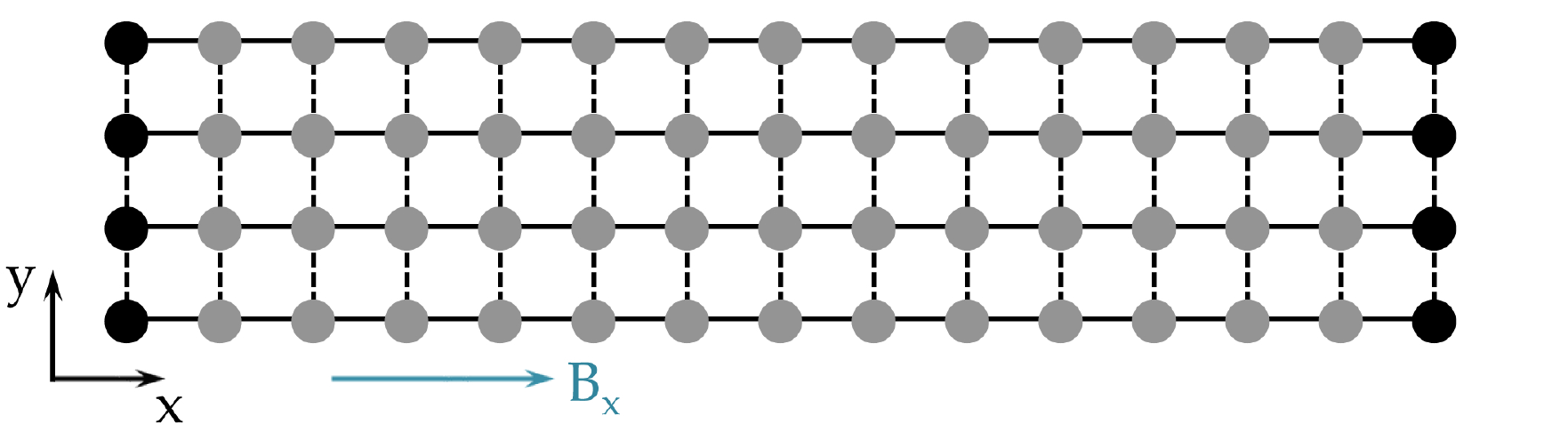}
	\caption{(color online) A sketch of a finite-size strip with a magnetic field $B=B_x$ along the $x$-axis. The black sites denote the short edges of the system where the Majorana modes would be localized. This system can be thought of as a set of 1D wires coupled in the $y$-direction.}
	\label{quasi-1Dsketch}
\end{figure}

In what follows we focus on finite-size strips, i.e. systems made-up of $N_y > 1$ coupled wires each with a finite but large number of sites $N_x \gg 1$ and $N_x \gg N_y$. We consider an in-plane magnetic field $B_x$ parallel to the long edge of the system (see Fig.~\ref{quasi-1Dsketch}). Note that for a magnetic field parallel to the $y$-axis, no Majorana states can form, since the magnetic field would in this case be parallel to the direction of the spin-orbit coupling in the wires. A similar system was considered in Ref.~[\onlinecite{Sedlmayr2016}], but only for magnetic fields perpendicular to the plane of the system. We apply open boundary conditions in both $x$ and $y$ directions. 

\begin{figure}[h!]
	\begin{tabular}{cc}
	{\textbf{Majorana polarisation}} & {\textbf{Singular points}$\phantom{aa}$} \\
	\includegraphics*[width = 0.45\columnwidth]{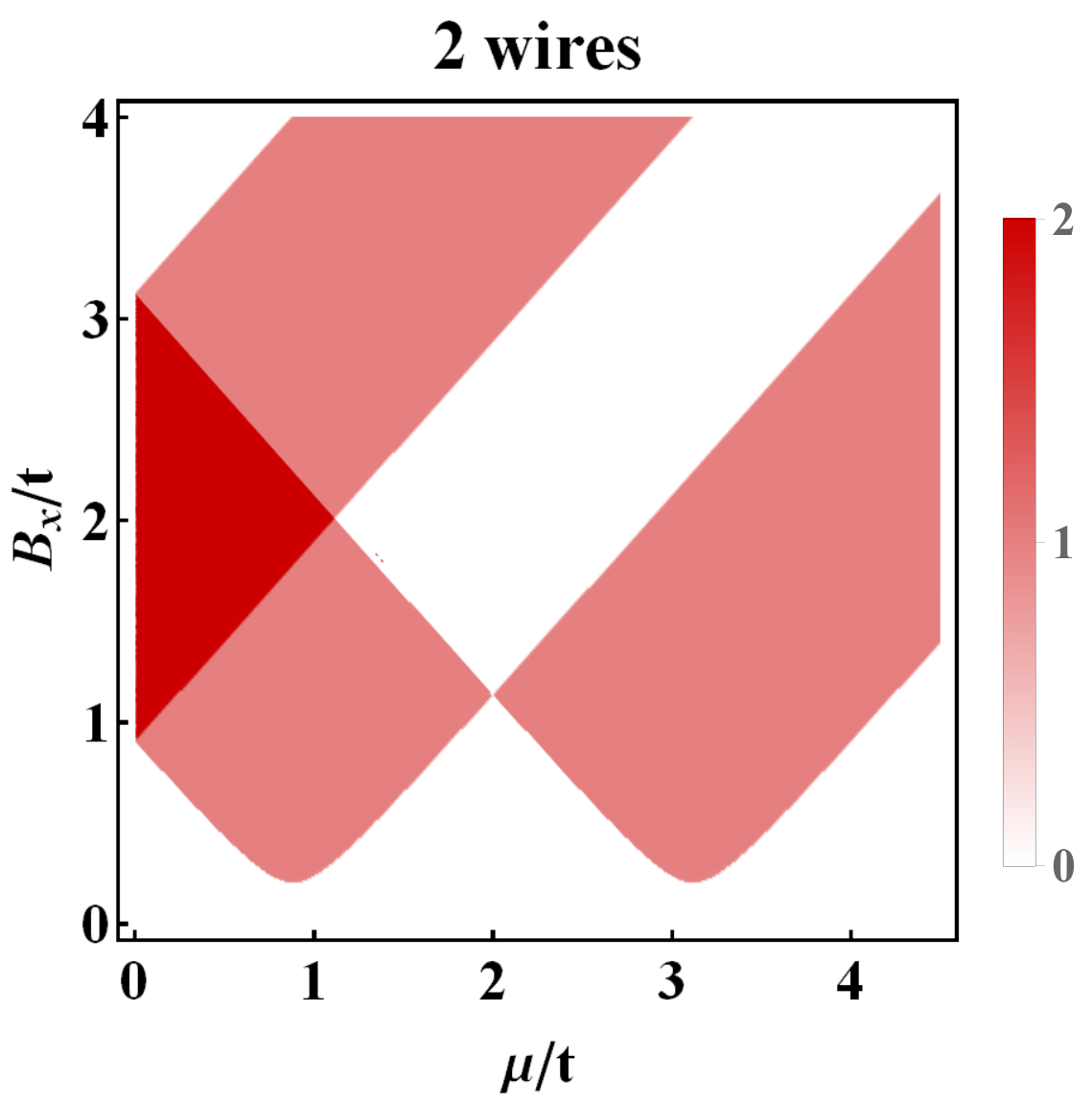} &
	\includegraphics*[width = 0.47\columnwidth]{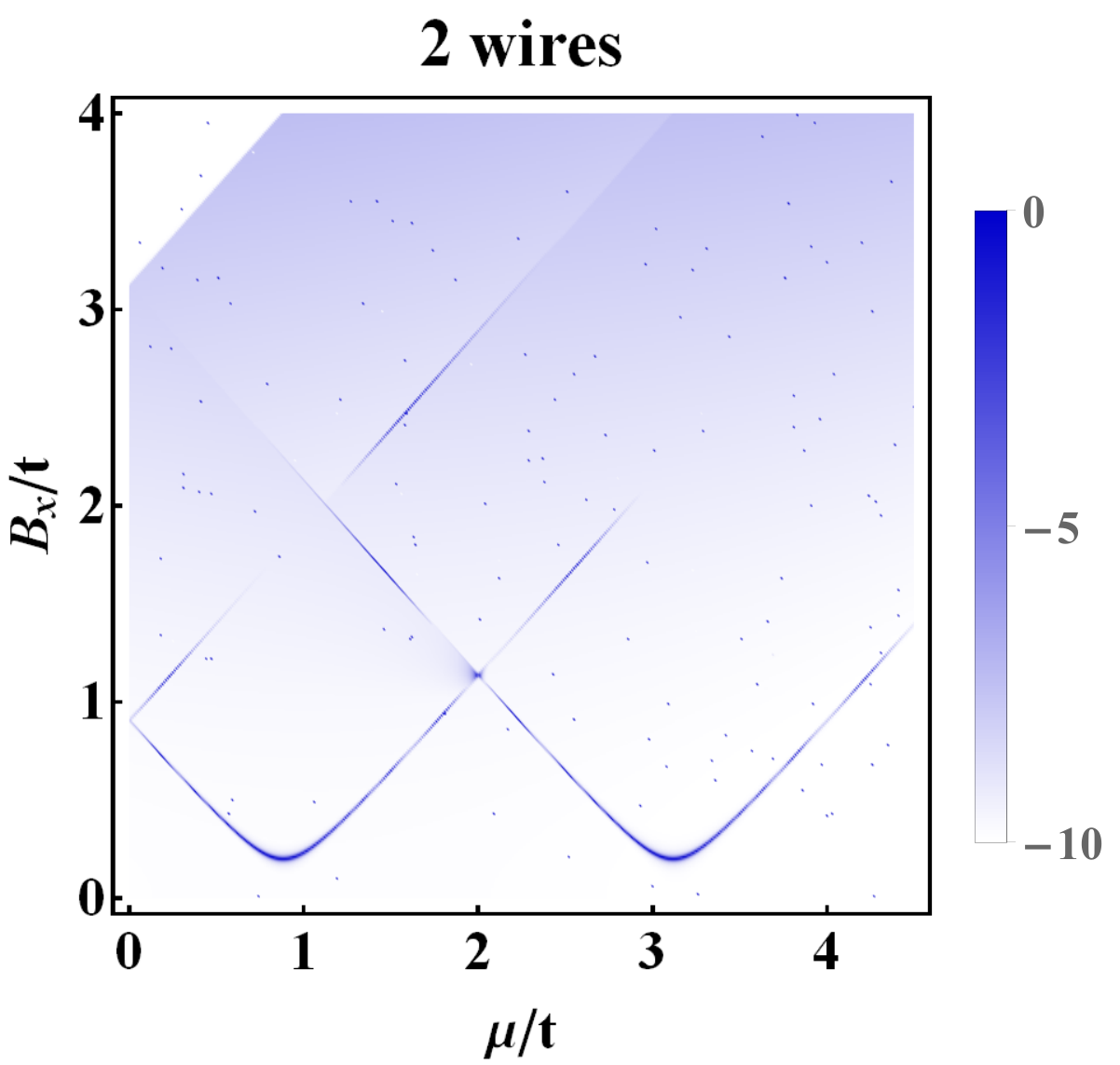}\\
	\includegraphics*[width = 0.45\columnwidth]{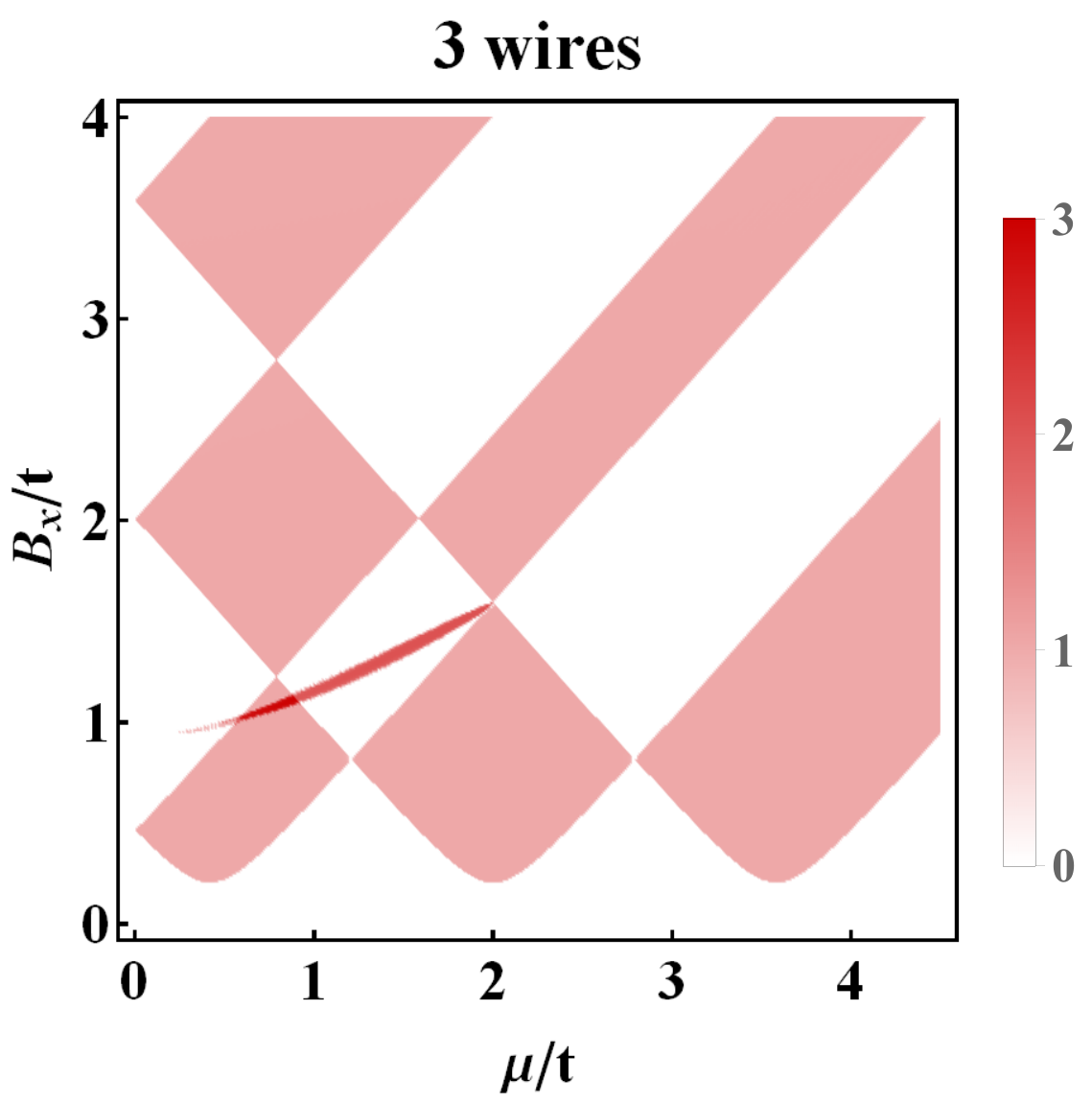} &
	\includegraphics*[width = 0.47\columnwidth]{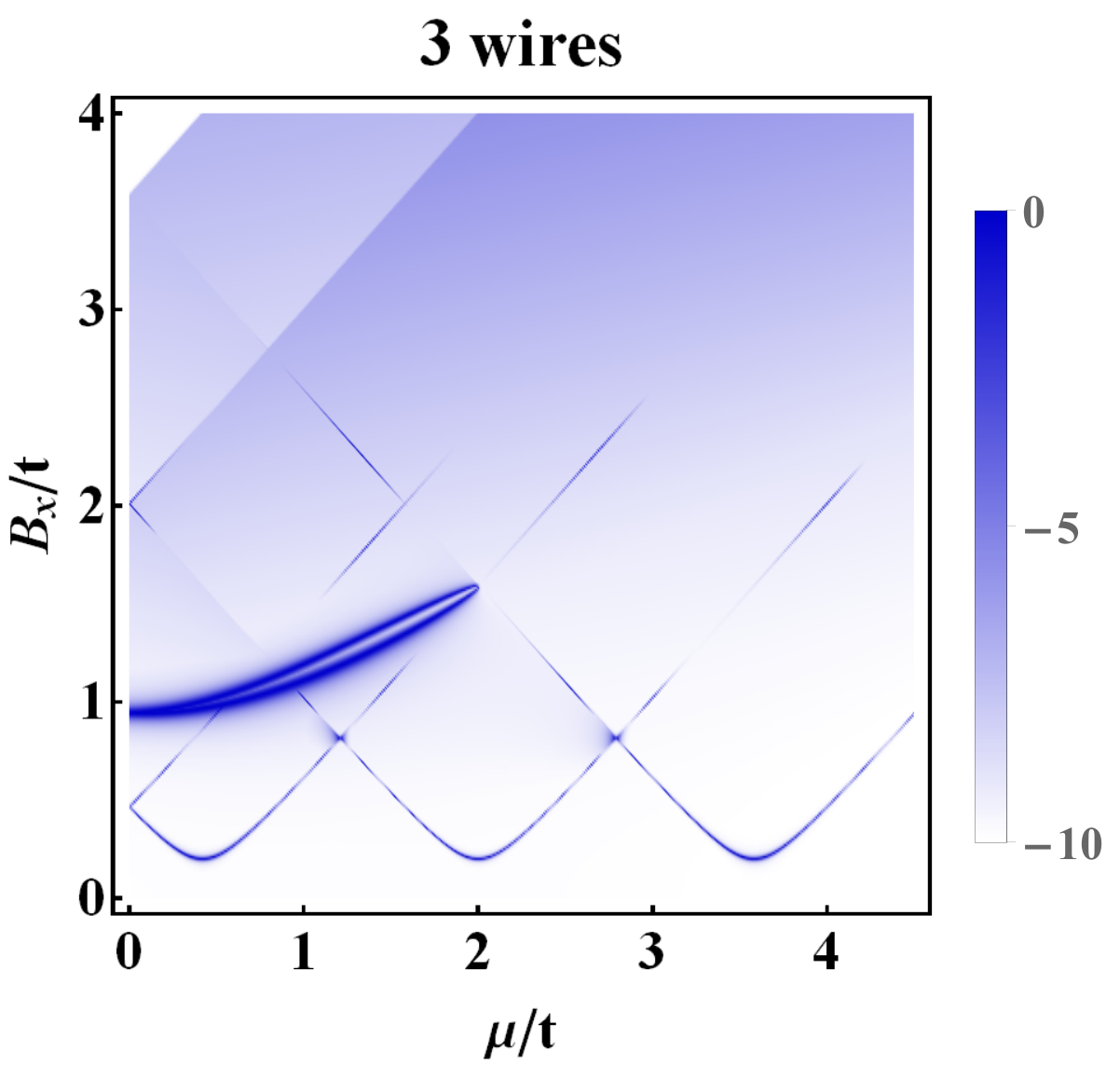}\\
	\includegraphics*[width = 0.45\columnwidth]{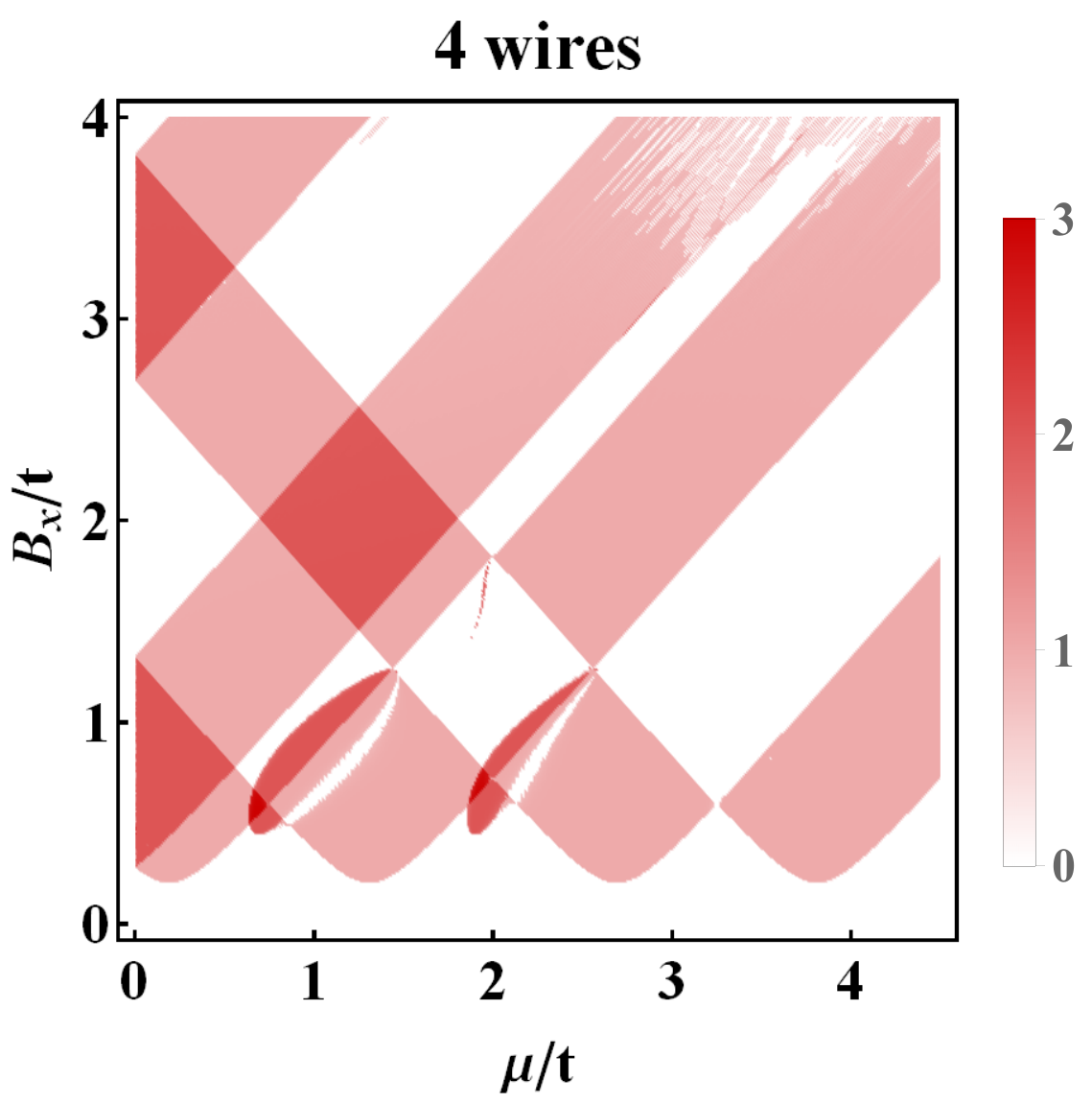} &
	\includegraphics*[width = 0.47\columnwidth]{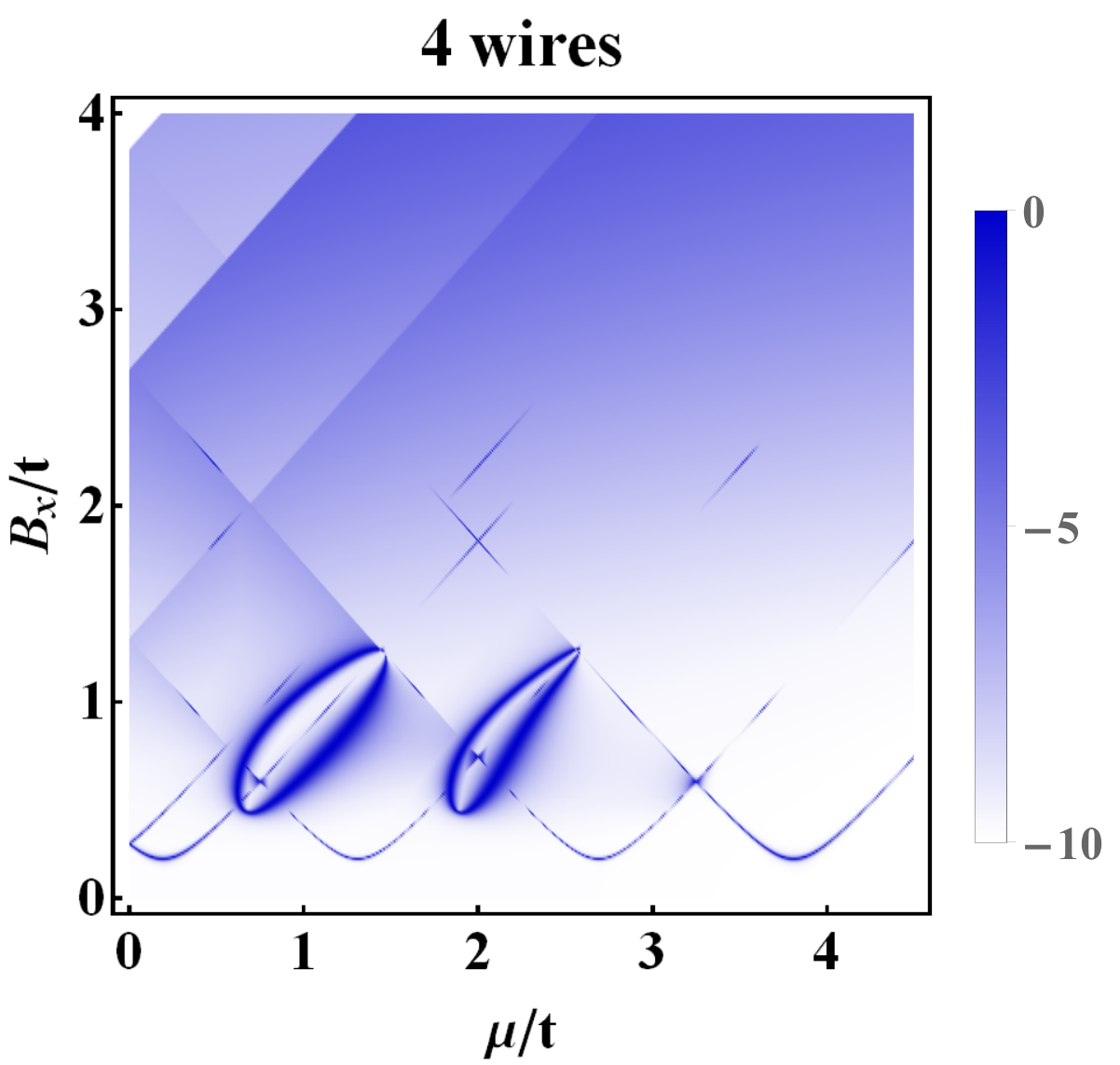}\\
	\end{tabular}
	\caption{(color online) Topological phase diagrams in the $(\mu, B_x)$ plane. In the left column we plot the total MP summed over all the low-energy states with a MP larger than a given cutoff, here taken to be $0.8$. The color scheme indicates the number of MBS pairs. In the right column we plot the results of the SP calculation; the phase transition lines are given by the zeroes of the plot. We consider finite-size strips with 2, 3 and 4 coupled wires. The results of these two methods are consistent, and the phase transition lines coincide. We set $\Delta=0.2t, \lambda_x=\lambda_y=0.5t$. }
	\label{234wires-PhD}
\end{figure}

As described in Section II, we will use two main tools to obtain the phase diagram for these systems. The first is to numerically diagonalize the tight-binding Hamiltonian and employ the integrated MP by plotting its value as a function of the chemical potential and the magnetic field. The second is to assume that the momentum $k_x$ along $x$ is a "good" quantum number, and exploit the SP technique. 
In Fig.~\ref{234wires-PhD} we show numerically that the results of the first two methods are fully consistent. Each phase transition boundary defined as a change in the number of MBS pairs obtained via the MP technique corresponds to a line of zeroes in the SP plot. The only apparent exception is the special case of the white lines in the $N_y=4$ close to $\mu=t$ and $\mu=2.2t$ and $B=0.5t$. They seem to correspond to the special case of a zero-size non-topological line-like region between two topological regions with one pair of MBS. Such situations, i.e. topological regions divided by a line of non-topological points, can arise, and are well captured here by the SP method. The numerical method is a bit less precise in this case, and the non-topological lines acquire a finite width, mostly because of the finite length of the considered systems (in the infinite-length limit the width of these regions should go to zero).

Note that the integrated MP allows not only to show the phase transition boundaries, but also to give access to the number of emerging MBS. Since the chiral symmetry (a combination of the PHS and the TRS) is absent, we cannot easily use the counting formula introduced in Ref.~[\onlinecite{Mandal2016b}] to obtain the number of Majorana modes using this method. The counting formula is in principle also applicable in the absence of the chiral symmetry, but the broken TRS case is very cumbersome and much harder to implement numerically. Therefore, we use here the SP technique only to obtain the phase transition lines; the actual number of the MBS pairs and the topological character of a given phase space region are obtained numerically via the calculation of the total MP. We should point out that some segments of the phase transition lines in the right column of Fig.~\ref{234wires-PhD} are almost non-visible. This is not due to the failure of the method, but to the numerical grid: the regions in which the zeroes of the determinant occur are very thin and the number of points required in the grid would be too large to capture them entirely. We did check though that the phase transition lines are present everywhere as expected, even if not fully shown in Fig.~\ref{234wires-PhD}.
 
Note also that by increasing the number of wires we can increase the number of MBS. However, as we will show in the next section only the states with an odd number of MBS pairs are topologically protected.

\section{Effects of disorder and topological invariant calculations}

\begin{figure}[h!]
	\centering
	\begin{tabular}{ccc}
		{\textbf{MP with disorder}}$\phantom{a}$ & {\textbf{TI without disorder}} \\
		\includegraphics*[width=0.47\columnwidth]{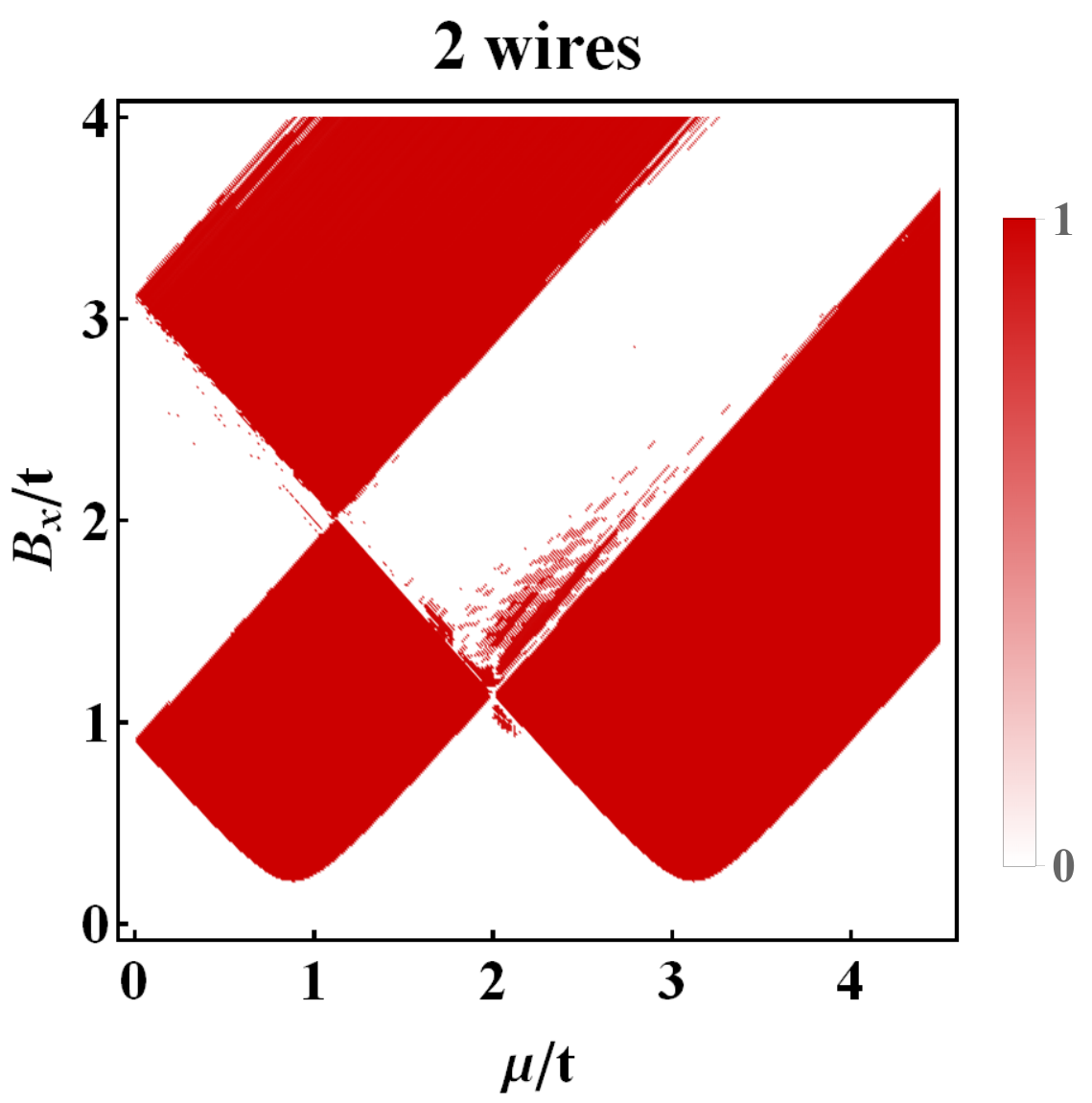} & 
		\includegraphics*[width=0.42\columnwidth]{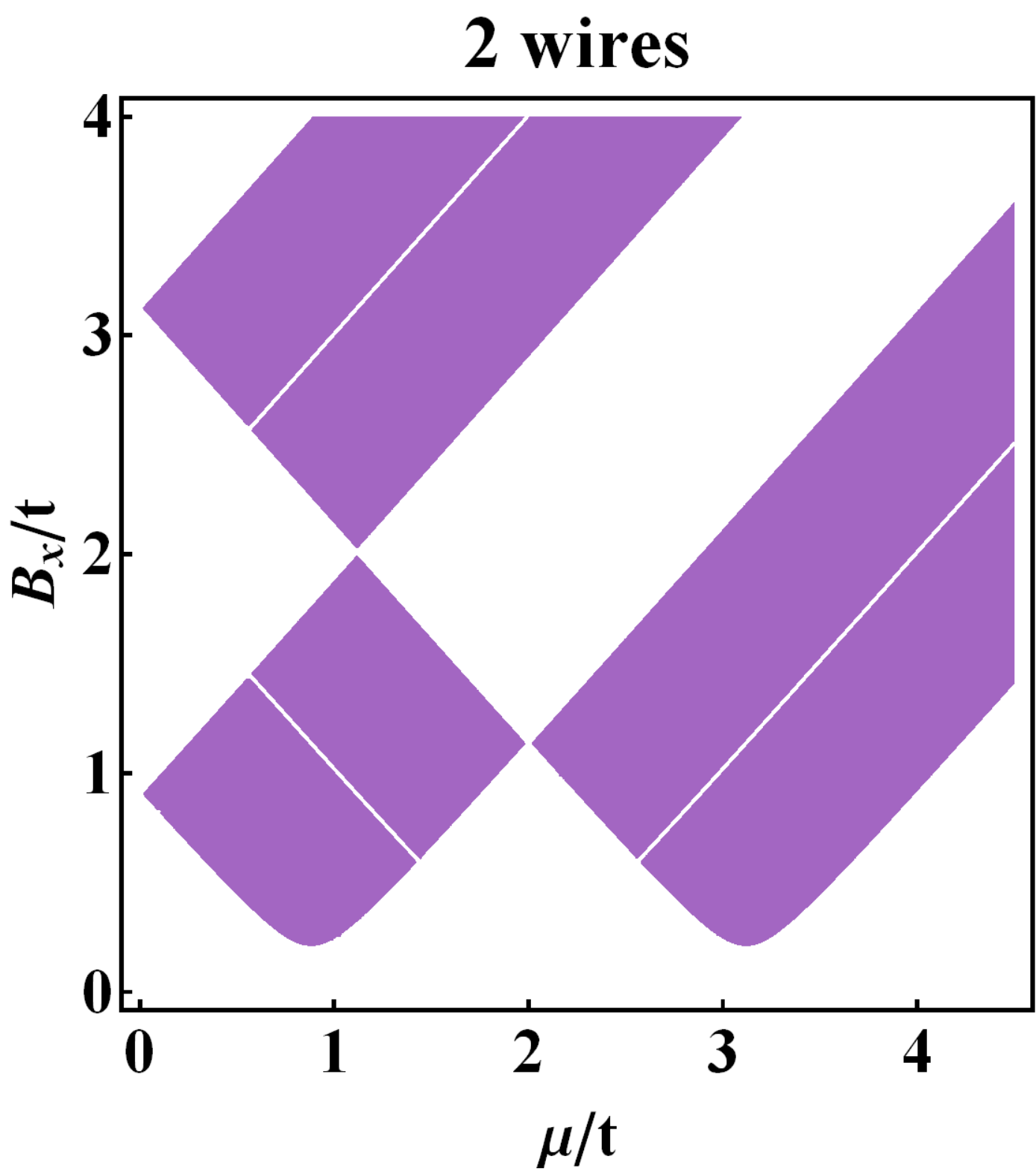}	\\	
		\includegraphics*[width=0.47\columnwidth]{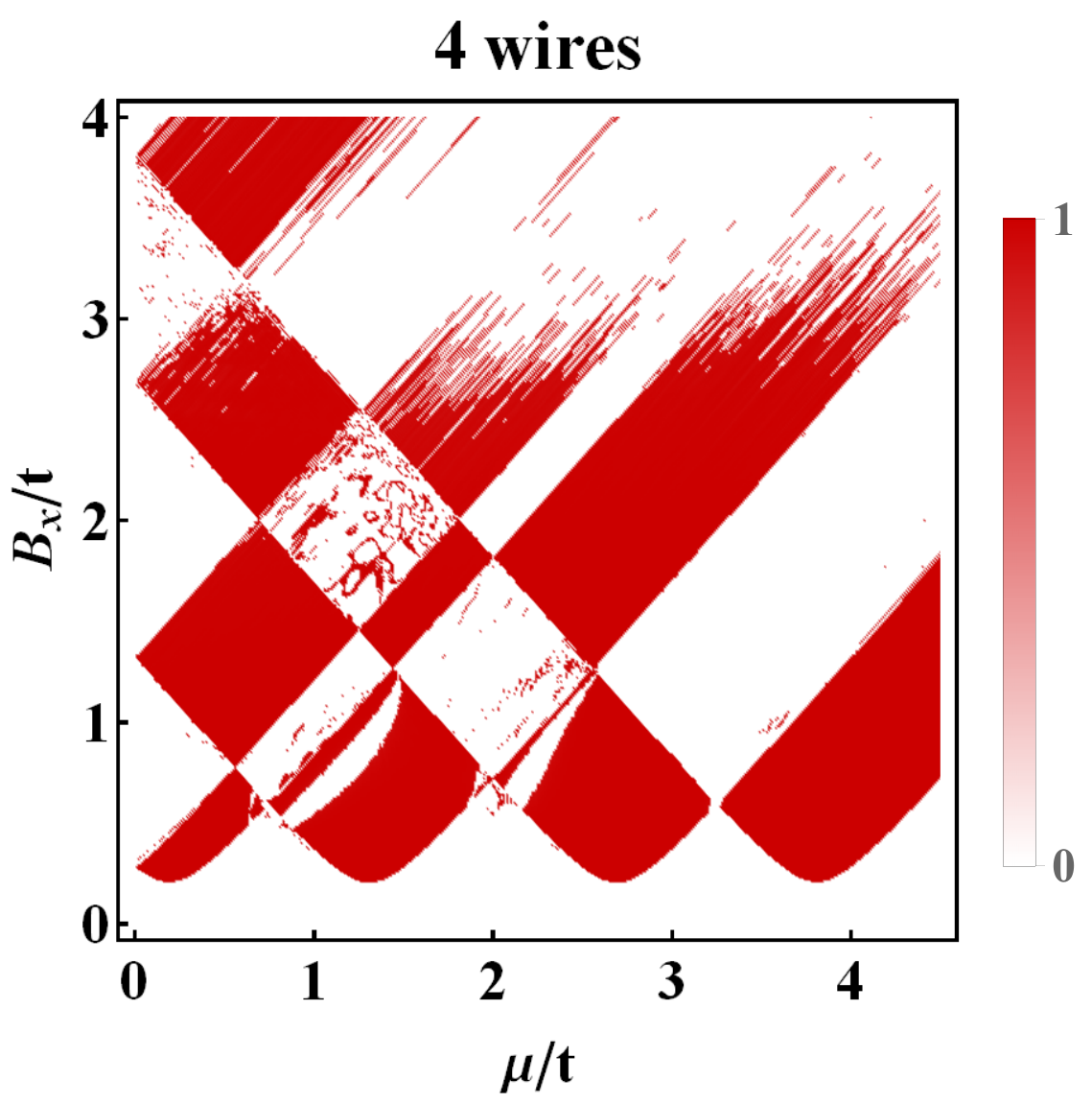} &
		\includegraphics*[width=0.42\columnwidth]{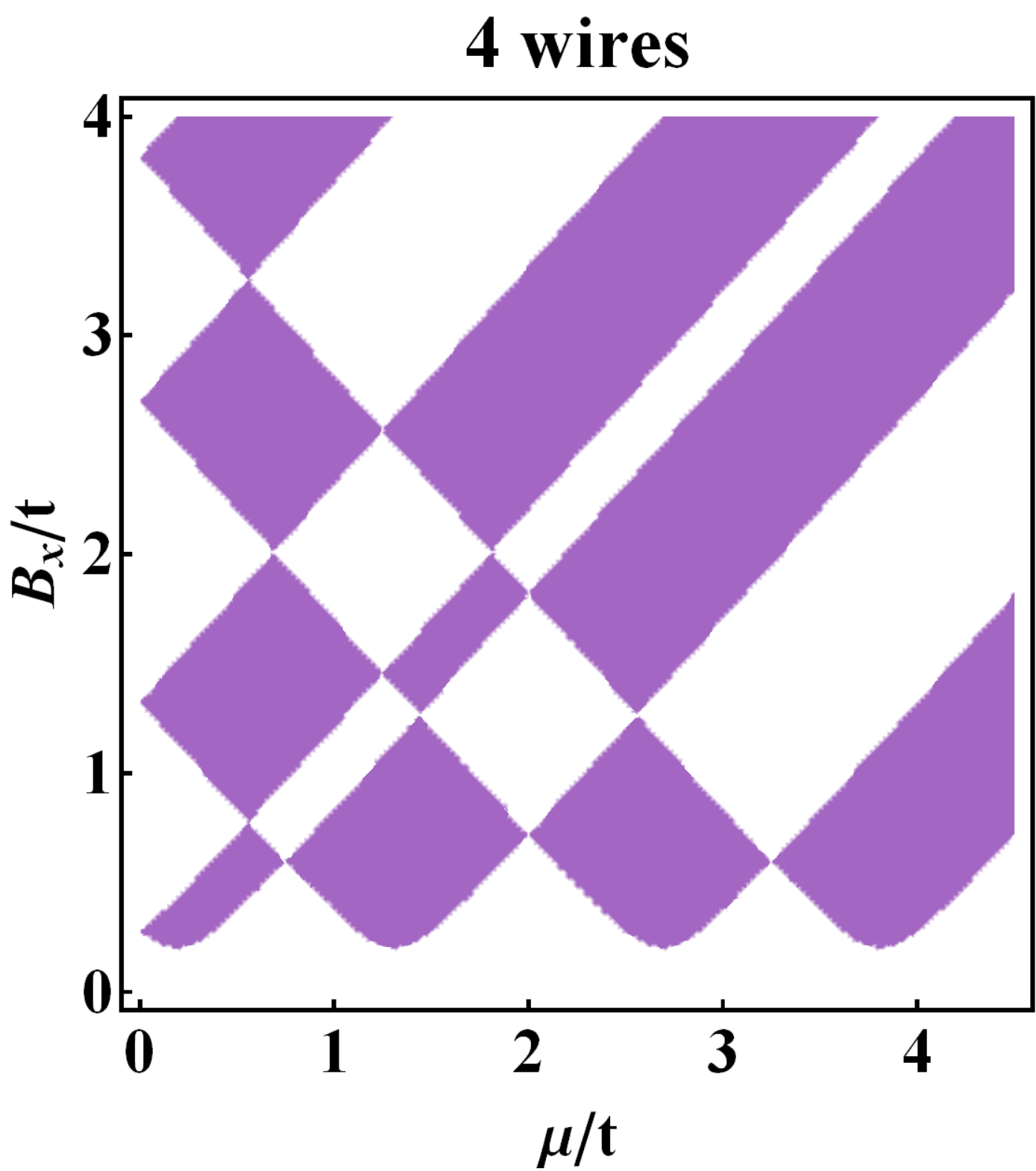} 
	\end{tabular}
	\caption{(color online) Topological phase diagrams in the $(\mu, B_x)$ plane for $2$ and $4$ coupled wires. In the left column we depict the MP of the lowest-energy mode for a disordered system. We impose a MP cutoff of $0.95$ (the states with MP smaller than $1$ cannot be considered actual Majoranas and usually correspond to non-zero energies, even if they remain Majorana-like). In the right column we depict the phase diagram as obtained using the topological invariant calculation (without disorder) (the topological regions are shown in violet). Note that, up to some special-points lines, the TI results are fully consistent with the those for the MP in the disordered system, which is expected since the TI gives access to the parity of the number of the MBS. In all the panels $\Delta=0.2t, \lambda_x=\lambda_y=0.5t$.}
		\label{disorderq1D}
\end{figure}

In what follows we show that a small amount of disorder makes the Majorana modes re-combine and form regular electronic states in all the regions in the parameter space with an even number of Majorana modes; thus these regions are not topologically protected. However, in all the regions with odd numbers of Majorana modes one MBS pair survives the effects of disorder. In the left column of Fig. \ref{disorderq1D} we show the phase diagrams for finite-size strips in the presence of disorder. The disorder considered here is a random variation of the value of the Zeeman magnetic field with an intensity of $5\%$ around its average value.\cite{Sedlmayr2015a} Indeed, we see that in the even-parity regions of the phase space the Majorana modes are destroyed, confirming their non-topological character.

We also present the corresponding phase diagrams computed using the topological invariant (TI) (for the details of the derivations see Appendix A, as well as Refs.~[\onlinecite{Sedlmayr2016}] and [\onlinecite{Gibertini2012}]). In Fig. \ref{disorderq1D}  we compare the phase diagrams showing the topological regions surviving the effects of disorder (left column) and those obtained using the topological invariant (right column). Indeed, up to some sets of lines of special points, the topological regions, as predicted by the topological invariant, coincide with the regions in the phase diagram shown numerically to exhibit an odd-parity of MBS pairs and to survive the presence of disorder.

\begin{figure}[h!]
	\centering
	\includegraphics*[width = 0.99\columnwidth]{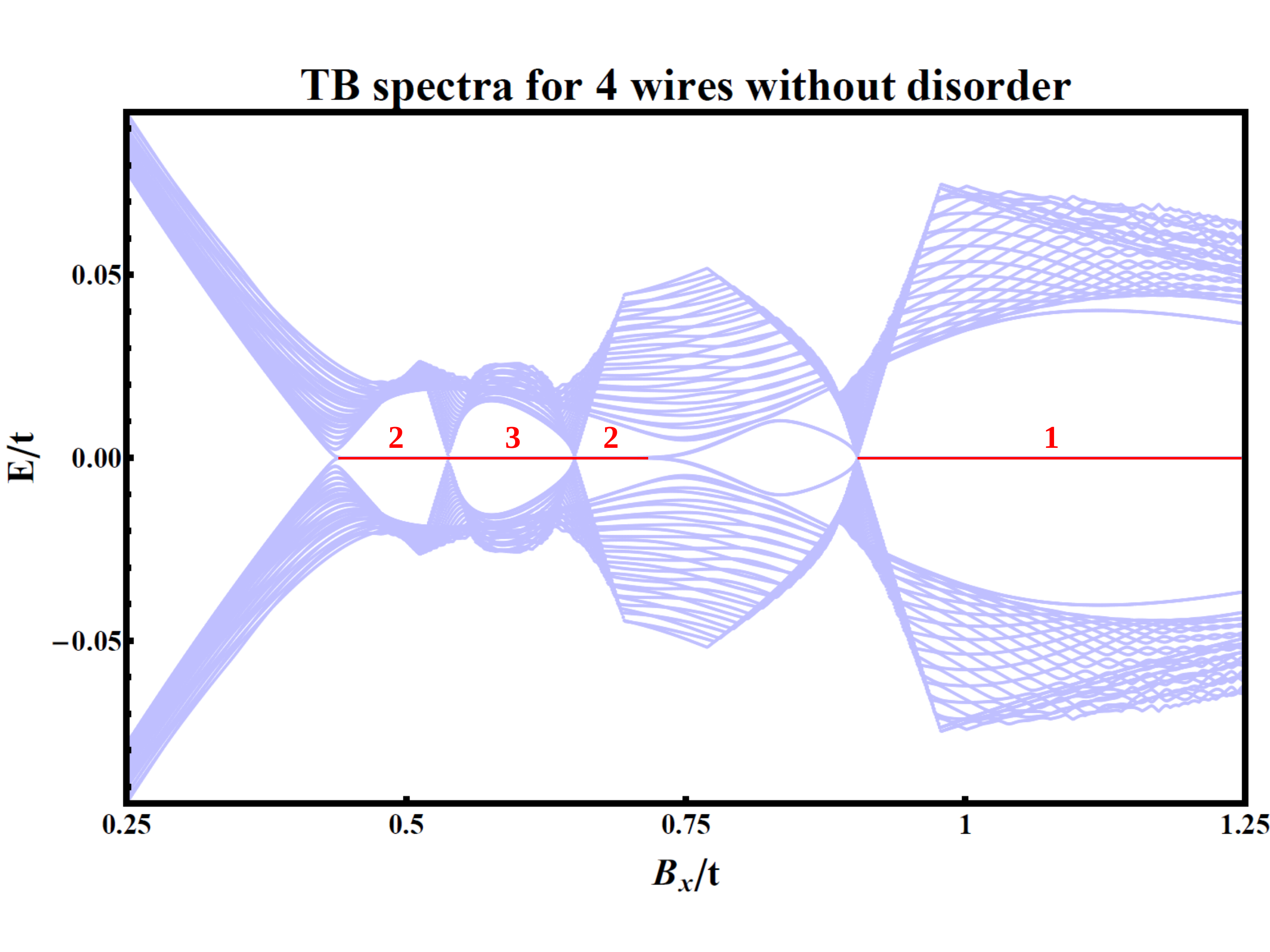}
	\includegraphics*[width = 0.99\columnwidth]{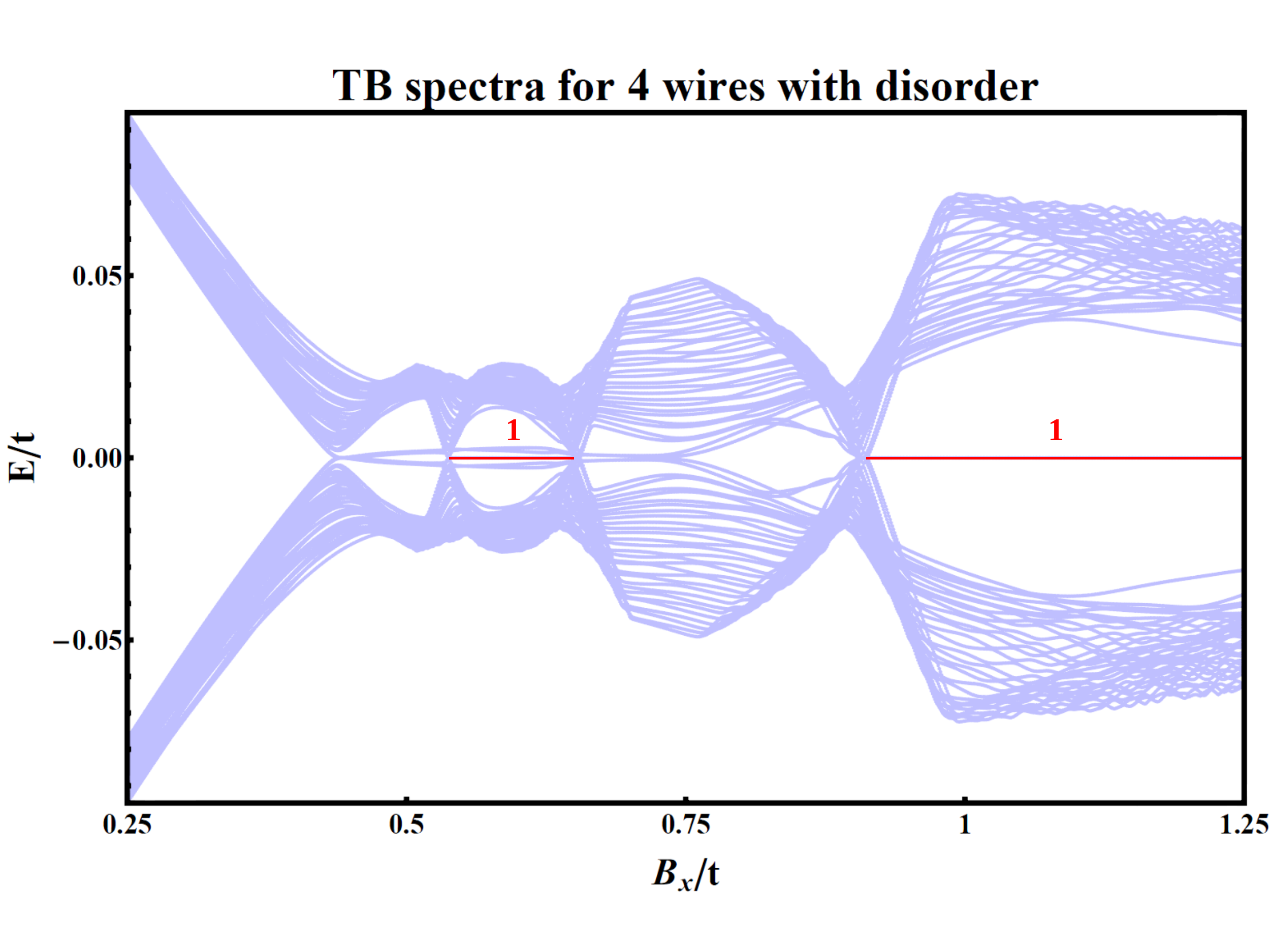}
	\caption{(color online) Energy spectra for 4-wires finite-size strips with and without disorder (lower and upper panel respectively) as a function of an in-plane magnetic field $B_x$, varying from $0.25t$ to $1.25t$. We restrict ourselves to plotting only the lowest 60 energy levels and we set $\Delta=0.2t, \lambda_x=\lambda_y=0.5t$, and $\mu=0.69t$. These panels correspond to vertical cuts of lower left panels in Fig.~\ref{234wires-PhD} and \ref{disorderq1D} respectively. The number of pairs is shown above the corresponding Majorana zero energy lines, highlighted in red. }
	\label{TBspectra300x4}
\end{figure}

It is also worth discussing how the value of the topological gap protecting the zero-energy states changes in the presence of disorder. In Fig.~\ref{TBspectra300x4} we plot the energy spectra of a 4-wire finite-size strip for a fixed value of the chemical potential as a function of an in-plane magnetic field $B_x$, both in the absence and in the presence of disorder. This corresponds to taking vertical cuts of the lower left panels of Figs.~\ref{234wires-PhD} and \ref{disorderq1D}. Without disorder, in full accordance with Fig.~\ref{234wires-PhD}, we have MBS for magnetic fields $B_x$ from $\sim0.45t$ to $\sim0.73t$ and $\sim0.92t$ to $\sim1.25t$. It is worth mentioning that certain regions contain more than one pair of MBS. The number of pairs is shown above the corresponding Majorana zero energy lines, highlighted in red. First, we note that, consistent with the phase diagrams presented in Figs.~\ref{234wires-PhD} and \ref{disorderq1D}, all the regions with odd numbers of MBS are protected against disorder, exhibiting one stable zero-energy mode (\emph{cf.} regions from $\sim0.55t$ to $\sim0.64t$ and $\sim0.92t$ to $\sim1.25t$ respectively), whereas in the regions with even numbers these states acquire a finite energy in the presence of disorder, confirming that these regions in the phase diagram are not topologically protected. Moreover we see that the gap protecting these zero-energy states is affected slightly by disorder, more significantly for the states with even numbers of Majoranas, consistent with the lack of topological protection for these states. 

\section{Finite-size squares}

If both $N_{x,y} \gg 1$ and are comparable in size then we are dealing with a finite-size square. In Ref.~[\onlinecite{Sedlmayr2016}] it has been shown that for perpendicular Zeeman magnetic fields finite-energy quasi-Majorana-like states ($C=\sqrt{2}/2$) may form, localized mostly in the corners of these square flakes, for a set of parameters inside the 2D bulk topological phase. However, for in-plane magnetic fields (e.g. along $x$-axis) the situation is very different since the rotation symmetry is broken and we always have an in-plane special direction, and we can no longer expect quasi-Majorana states with rotationally symmetric MP. 

\begin{figure}
	\centering
	\vspace{.2in}
	\includegraphics*[width = 0.6\columnwidth]{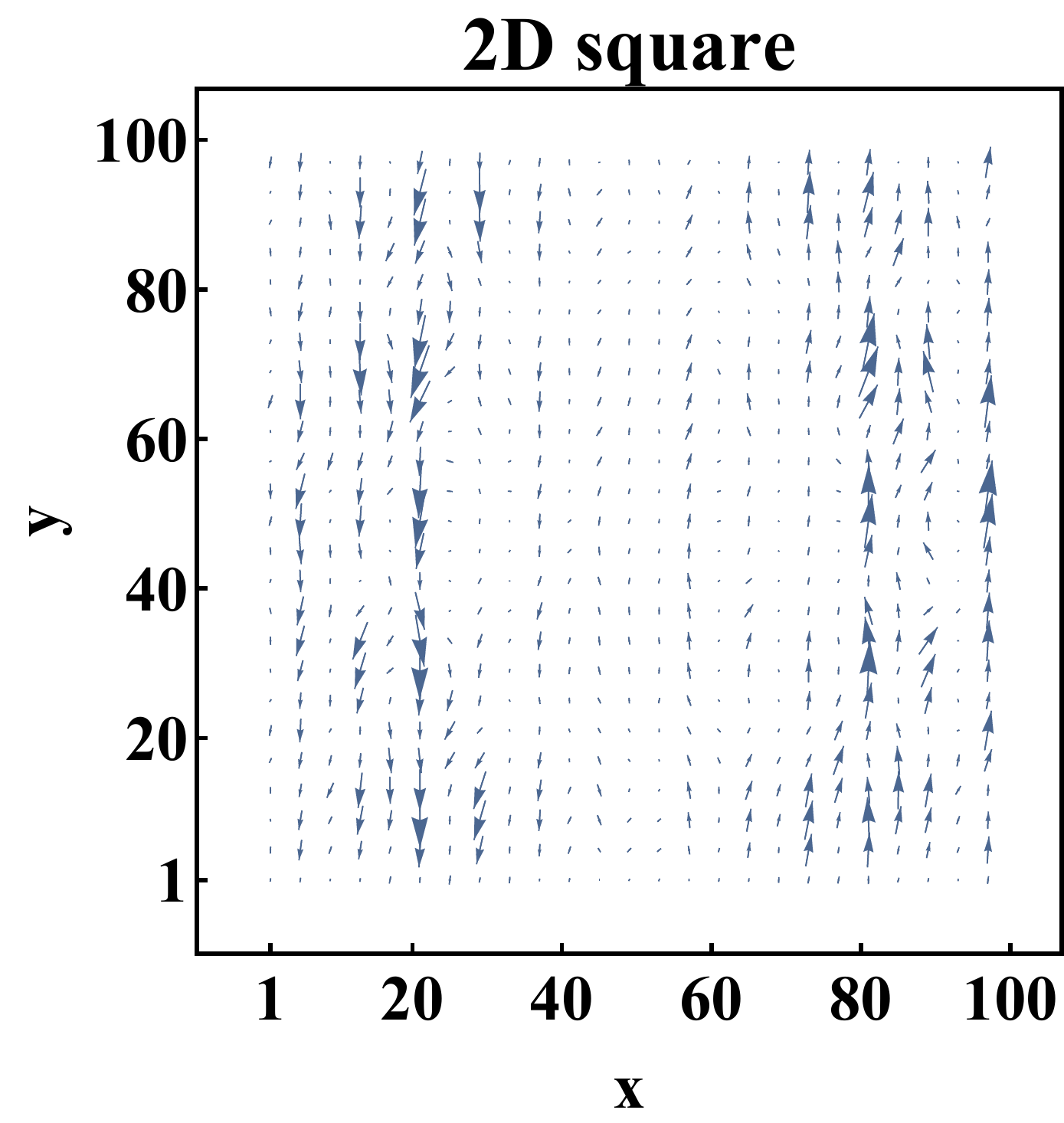}
	\caption{(color online) The MP vector for a finite-size square consisting of $100\times 100$ sites in a magnetic field along the $x$-axis. We choose a set of parameters $\mu=4t, B_x=0.28t, \Delta=0.2t, \lambda_x=\lambda_y=0.5t$.}
	\label{squareMPvector}
\end{figure}

By analyzing the MP of these systems we note that the generic situation that emerges is that depicted in Fig.~\ref{squareMPvector}: quais-disordered edge states localized on the edges of the system perpendicular to the magnetic field. Such states have also a quasi-disordered MP, and the integral of the MP over one of these edges states is finite (for the case in  Fig.~\ref{squareMPvector} this is of the order of  $0.9$).

This tendency to form a Majorana state is larger for values of the magnetic field close to the transition, and for systems with a very large $N_x$ we can actually recover actual Majorana states in these systems on the edges perpendicular to the magnetic field. The systems required to recover a full Majorana are too large for our numerical abilities, but even for smaller systems we have managed to tune up the parameters to get a MP up to $0.9$, and increasing the size will improve this value. This is important from an experimental perspective, since it indicates that for in-plane fields actual Majorana states can form even in wide square systems, while for perpendicular fields this can never be the case unless one dimension is much larger than the other one.\\

\section{Conclusions} \label{Conclusions}

We have studied the formation of Majorana bound states in infinite ribbons, finite-size strips and squares with Rashba spin-orbit coupling and an in-plane magnetic field. We have shown that in infinite ribbons chiral Majorana fermions may form when the magnetic field is perpendicular to the edges of the ribbon. Furthermore, we have studied finite-size strips exploiting a numerical diagonalization technique and the Majorana polarization, as well as the singular points technique, and we have proven the qualitative equivalence of these two methods in constructing the phase diagrams of these systems. We have also evaluated the topological invariant for the finite-size strips and we have shown that its usage allows one to obtain the correct phase diagrams for the parity of the number of MBS pairs. Moreover, we have confirmed numerically that the phases with even number of MBS pairs are not stable in the presence of disorder, and are thus topologically trivial, while the phases with an odd number of MBS preserve their topological character.

\section{Acknowledgements}

We would like to thank Nicholas Sedlmayr for fruitful discussions and useful comments on our manuscript. VK is grateful for the hospitality provided by the Perimeter Institute where a part of this work was performed. The research at the Perimeter Institute was supported in part by the Government of Canada through Industry Canada and by the Province of Ontario through the Ministry of Research and Information. The research at IPhT was supported in part by the ERC Starting Independent Researcher Grant NANOGRAPHENE 256965.\\

\section{Author contribution statement}

Cristina Bena suggested the issue, Ipsita Mandal helped with implementing the singular points technique, and Vardan Kaladzhyan, Julien Despres and Cristina Bena carried out the calculations. All the authors wrote and revised this article.

\bibliography{biblio_MF}

\begin{thebibliography}{57}%
\makeatletter
\providecommand \@ifxundefined [1]{%
 \@ifx{#1\undefined}
}%
\providecommand \@ifnum [1]{%
 \ifnum #1\expandafter \@firstoftwo
 \else \expandafter \@secondoftwo
 \fi
}%
\providecommand \@ifx [1]{%
 \ifx #1\expandafter \@firstoftwo
 \else \expandafter \@secondoftwo
 \fi
}%
\providecommand \natexlab [1]{#1}%
\providecommand \enquote  [1]{``#1''}%
\providecommand \bibnamefont  [1]{#1}%
\providecommand \bibfnamefont [1]{#1}%
\providecommand \citenamefont [1]{#1}%
\providecommand \href@noop [0]{\@secondoftwo}%
\providecommand \href [0]{\begingroup \@sanitize@url \@href}%
\providecommand \@href[1]{\@@startlink{#1}\@@href}%
\providecommand \@@href[1]{\endgroup#1\@@endlink}%
\providecommand \@sanitize@url [0]{\catcode `\\12\catcode `\$12\catcode
  `\&12\catcode `\#12\catcode `\^12\catcode `\_12\catcode `\%12\relax}%
\providecommand \@@startlink[1]{}%
\providecommand \@@endlink[0]{}%
\providecommand \url  [0]{\begingroup\@sanitize@url \@url }%
\providecommand \@url [1]{\endgroup\@href {#1}{\urlprefix }}%
\providecommand \urlprefix  [0]{URL }%
\providecommand \Eprint [0]{\href }%
\providecommand \doibase [0]{http://dx.doi.org/}%
\providecommand \selectlanguage [0]{\@gobble}%
\providecommand \bibinfo  [0]{\@secondoftwo}%
\providecommand \bibfield  [0]{\@secondoftwo}%
\providecommand \translation [1]{[#1]}%
\providecommand \BibitemOpen [0]{}%
\providecommand \bibitemStop [0]{}%
\providecommand \bibitemNoStop [0]{.\EOS\space}%
\providecommand \EOS [0]{\spacefactor3000\relax}%
\providecommand \BibitemShut  [1]{\csname bibitem#1\endcsname}%
\let\auto@bib@innerbib\@empty
\bibitem [{\citenamefont {Mandal}(2015)}]{Mandal2015}%
  \BibitemOpen
  \bibfield  {author} {\bibinfo {author} {\bibfnamefont {I.}~\bibnamefont
  {Mandal}},\ }\href {\doibase http://dx.doi.org/10.1209/0295-5075/110/67005}
  {\bibfield  {journal} {\bibinfo  {journal} {Europhysics Letters}\ }\textbf
  {\bibinfo {volume} {110}},\ \bibinfo {pages} {67005} (\bibinfo {year}
  {2015})}\BibitemShut {NoStop}%
\bibitem [{\citenamefont {Mandal}\ and\ \citenamefont
  {Tewari}(2016)}]{Mandal2016a}%
  \BibitemOpen
  \bibfield  {author} {\bibinfo {author} {\bibfnamefont {I.}~\bibnamefont
  {Mandal}}\ and\ \bibinfo {author} {\bibfnamefont {S.}~\bibnamefont
  {Tewari}},\ }\href {\doibase http://dx.doi.org/10.1016/j.physe.2015.12.009}
  {\bibfield  {journal} {\bibinfo  {journal} {Physica E}\ }\textbf {\bibinfo
  {volume} {79}},\ \bibinfo {pages} {180 } (\bibinfo {year}
  {2016})}\BibitemShut {NoStop}%
\bibitem [{\citenamefont {Mandal}(2016)}]{Mandal2016b}%
  \BibitemOpen
  \bibfield  {author} {\bibinfo {author} {\bibfnamefont {I.}~\bibnamefont
  {Mandal}},\ }\href {\doibase http://dx.doi.org/10.5488/CMP.19.33703}
  {\bibfield  {journal} {\bibinfo  {journal} {Condensed Matter Physics}\
  }\textbf {\bibinfo {volume} {19}},\ \bibinfo {pages} {33703} (\bibinfo {year}
  {2016})}\BibitemShut {NoStop}%
\bibitem [{\citenamefont {San-Jose}\ \emph {et~al.}(2016)\citenamefont
  {San-Jose}, \citenamefont {Prada},\ and\ \citenamefont
  {Aguado}}]{Aguado2016}%
  \BibitemOpen
  \bibfield  {author} {\bibinfo {author} {\bibfnamefont {P.}~\bibnamefont
  {San-Jose}}, \bibinfo {author} {\bibfnamefont {E.}~\bibnamefont {Prada}}, \
  and\ \bibinfo {author} {\bibfnamefont {R.}~\bibnamefont {Aguado}},\ }\href
  {\doibase http://dx.doi.org/10.1038/srep21427} {\bibfield  {journal}
  {\bibinfo  {journal} {Scientific Reports}\ }\textbf {\bibinfo {volume} {6}},\
  \bibinfo {pages} {21427} (\bibinfo {year} {2016})}\BibitemShut {NoStop}%
\bibitem [{\citenamefont {Kitaev}(2001)}]{Kitaev2001}%
  \BibitemOpen
  \bibfield  {author} {\bibinfo {author} {\bibfnamefont {A.~Y.}\ \bibnamefont
  {Kitaev}},\ }\href {http://stacks.iop.org/1063-7869/44/i=10S/a=S29}
  {\bibfield  {journal} {\bibinfo  {journal} {Physics-Uspekhi}\ }\textbf
  {\bibinfo {volume} {44}},\ \bibinfo {pages} {131} (\bibinfo {year}
  {2001})}\BibitemShut {NoStop}%
\bibitem [{\citenamefont {Fu}\ and\ \citenamefont {Kane}(2008)}]{Fu2008}%
  \BibitemOpen
  \bibfield  {author} {\bibinfo {author} {\bibfnamefont {L.}~\bibnamefont
  {Fu}}\ and\ \bibinfo {author} {\bibfnamefont {C.~L.}\ \bibnamefont {Kane}},\
  }\href {\doibase 10.1103/PhysRevLett.100.096407} {\bibfield  {journal}
  {\bibinfo  {journal} {Phys. Rev. Lett.}\ }\textbf {\bibinfo {volume} {100}},\
  \bibinfo {pages} {096407} (\bibinfo {year} {2008})}\BibitemShut {NoStop}%
\bibitem [{\citenamefont {Fu}\ and\ \citenamefont {Kane}(2009)}]{Fu2009}%
  \BibitemOpen
  \bibfield  {author} {\bibinfo {author} {\bibfnamefont {L.}~\bibnamefont
  {Fu}}\ and\ \bibinfo {author} {\bibfnamefont {C.~L.}\ \bibnamefont {Kane}},\
  }\href {\doibase 10.1103/PhysRevLett.102.216403} {\bibfield  {journal}
  {\bibinfo  {journal} {Phys. Rev. Lett.}\ }\textbf {\bibinfo {volume} {102}},\
  \bibinfo {pages} {216403} (\bibinfo {year} {2009})}\BibitemShut {NoStop}%
\bibitem [{\citenamefont {Sato}\ \emph {et~al.}(2009)\citenamefont {Sato},
  \citenamefont {Takahashi},\ and\ \citenamefont {Fujimoto}}]{Sato2009}%
  \BibitemOpen
  \bibfield  {author} {\bibinfo {author} {\bibfnamefont {M.}~\bibnamefont
  {Sato}}, \bibinfo {author} {\bibfnamefont {Y.}~\bibnamefont {Takahashi}}, \
  and\ \bibinfo {author} {\bibfnamefont {S.}~\bibnamefont {Fujimoto}},\ }\href
  {\doibase 10.1103/PhysRevLett.103.020401} {\bibfield  {journal} {\bibinfo
  {journal} {Phys. Rev. Lett.}\ }\textbf {\bibinfo {volume} {103}},\ \bibinfo
  {pages} {020401} (\bibinfo {year} {2009})}\BibitemShut {NoStop}%
\bibitem [{\citenamefont {Lutchyn}\ \emph {et~al.}(2010)\citenamefont
  {Lutchyn}, \citenamefont {Sau},\ and\ \citenamefont
  {Das~Sarma}}]{Lutchyn2010}%
  \BibitemOpen
  \bibfield  {author} {\bibinfo {author} {\bibfnamefont {R.~M.}\ \bibnamefont
  {Lutchyn}}, \bibinfo {author} {\bibfnamefont {J.~D.}\ \bibnamefont {Sau}}, \
  and\ \bibinfo {author} {\bibfnamefont {S.}~\bibnamefont {Das~Sarma}},\ }\href
  {\doibase 10.1103/PhysRevLett.105.077001} {\bibfield  {journal} {\bibinfo
  {journal} {Phys. Rev. Lett.}\ }\textbf {\bibinfo {volume} {105}},\ \bibinfo
  {pages} {077001} (\bibinfo {year} {2010})}\BibitemShut {NoStop}%
\bibitem [{\citenamefont {Oreg}\ \emph {et~al.}(2010)\citenamefont {Oreg},
  \citenamefont {Refael},\ and\ \citenamefont {von Oppen}}]{Oreg2010}%
  \BibitemOpen
  \bibfield  {author} {\bibinfo {author} {\bibfnamefont {Y.}~\bibnamefont
  {Oreg}}, \bibinfo {author} {\bibfnamefont {G.}~\bibnamefont {Refael}}, \ and\
  \bibinfo {author} {\bibfnamefont {F.}~\bibnamefont {von Oppen}},\ }\href
  {\doibase 10.1103/PhysRevLett.105.177002} {\bibfield  {journal} {\bibinfo
  {journal} {Phys. Rev. Lett.}\ }\textbf {\bibinfo {volume} {105}},\ \bibinfo
  {pages} {177002} (\bibinfo {year} {2010})}\BibitemShut {NoStop}%
\bibitem [{\citenamefont {Potter}\ and\ \citenamefont
  {Lee}(2010)}]{Potter2010}%
  \BibitemOpen
  \bibfield  {author} {\bibinfo {author} {\bibfnamefont {A.~C.}\ \bibnamefont
  {Potter}}\ and\ \bibinfo {author} {\bibfnamefont {P.~A.}\ \bibnamefont
  {Lee}},\ }\href {\doibase 10.1103/PhysRevLett.105.227003} {\bibfield
  {journal} {\bibinfo  {journal} {Phys. Rev. Lett.}\ }\textbf {\bibinfo
  {volume} {105}},\ \bibinfo {pages} {227003} (\bibinfo {year}
  {2010})}\BibitemShut {NoStop}%
\bibitem [{\citenamefont {Qi}\ and\ \citenamefont {Zhang}(2011)}]{Qi2011}%
  \BibitemOpen
  \bibfield  {author} {\bibinfo {author} {\bibfnamefont {X.-L.}\ \bibnamefont
  {Qi}}\ and\ \bibinfo {author} {\bibfnamefont {S.-C.}\ \bibnamefont {Zhang}},\
  }\href {\doibase 10.1103/RevModPhys.83.1057} {\bibfield  {journal} {\bibinfo
  {journal} {Rev. Mod. Phys.}\ }\textbf {\bibinfo {volume} {83}},\ \bibinfo
  {pages} {1057} (\bibinfo {year} {2011})}\BibitemShut {NoStop}%
\bibitem [{\citenamefont {Martin}\ and\ \citenamefont
  {Morpurgo}(2012)}]{Martin2012}%
  \BibitemOpen
  \bibfield  {author} {\bibinfo {author} {\bibfnamefont {I.}~\bibnamefont
  {Martin}}\ and\ \bibinfo {author} {\bibfnamefont {A.~F.}\ \bibnamefont
  {Morpurgo}},\ }\href {\doibase 10.1103/PhysRevB.85.144505} {\bibfield
  {journal} {\bibinfo  {journal} {Phys. Rev. B}\ }\textbf {\bibinfo {volume}
  {85}},\ \bibinfo {pages} {144505} (\bibinfo {year} {2012})}\BibitemShut
  {NoStop}%
\bibitem [{\citenamefont {Tewari}\ and\ \citenamefont
  {Sau}(2012)}]{Tewari2012}%
  \BibitemOpen
  \bibfield  {author} {\bibinfo {author} {\bibfnamefont {S.}~\bibnamefont
  {Tewari}}\ and\ \bibinfo {author} {\bibfnamefont {J.~D.}\ \bibnamefont
  {Sau}},\ }\href {\doibase 10.1103/PhysRevLett.109.150408} {\bibfield
  {journal} {\bibinfo  {journal} {Phys. Rev. Lett.}\ }\textbf {\bibinfo
  {volume} {109}},\ \bibinfo {pages} {150408} (\bibinfo {year}
  {2012})}\BibitemShut {NoStop}%
\bibitem [{\citenamefont {Klinovaja}\ and\ \citenamefont
  {Loss}(2012)}]{Klinovaja2012}%
  \BibitemOpen
  \bibfield  {author} {\bibinfo {author} {\bibfnamefont {J.}~\bibnamefont
  {Klinovaja}}\ and\ \bibinfo {author} {\bibfnamefont {D.}~\bibnamefont
  {Loss}},\ }\href {\doibase 10.1103/PhysRevB.86.085408} {\bibfield  {journal}
  {\bibinfo  {journal} {Phys. Rev. B}\ }\textbf {\bibinfo {volume} {86}},\
  \bibinfo {pages} {085408} (\bibinfo {year} {2012})}\BibitemShut {NoStop}%
\bibitem [{\citenamefont {Klinovaja}\ and\ \citenamefont
  {Loss}(2013)}]{Klinovaja2013}%
  \BibitemOpen
  \bibfield  {author} {\bibinfo {author} {\bibfnamefont {J.}~\bibnamefont
  {Klinovaja}}\ and\ \bibinfo {author} {\bibfnamefont {D.}~\bibnamefont
  {Loss}},\ }\href {\doibase 10.1103/PhysRevX.3.011008} {\bibfield  {journal}
  {\bibinfo  {journal} {Phys. Rev. X}\ }\textbf {\bibinfo {volume} {3}},\
  \bibinfo {pages} {011008} (\bibinfo {year} {2013})}\BibitemShut {NoStop}%
\bibitem [{\citenamefont {Nadj-Perge}\ \emph {et~al.}(2013)\citenamefont
  {Nadj-Perge}, \citenamefont {Drozdov}, \citenamefont {Bernevig},\ and\
  \citenamefont {Yazdani}}]{NadjPerge2013}%
  \BibitemOpen
  \bibfield  {author} {\bibinfo {author} {\bibfnamefont {S.}~\bibnamefont
  {Nadj-Perge}}, \bibinfo {author} {\bibfnamefont {I.~K.}\ \bibnamefont
  {Drozdov}}, \bibinfo {author} {\bibfnamefont {B.~A.}\ \bibnamefont
  {Bernevig}}, \ and\ \bibinfo {author} {\bibfnamefont {A.}~\bibnamefont
  {Yazdani}},\ }\href {\doibase 10.1103/PhysRevB.88.020407} {\bibfield
  {journal} {\bibinfo  {journal} {Phys. Rev. B}\ }\textbf {\bibinfo {volume}
  {88}},\ \bibinfo {pages} {020407} (\bibinfo {year} {2013})}\BibitemShut
  {NoStop}%
\bibitem [{\citenamefont {Sau}\ and\ \citenamefont {Demler}(2013)}]{Sau2013}%
  \BibitemOpen
  \bibfield  {author} {\bibinfo {author} {\bibfnamefont {J.~D.}\ \bibnamefont
  {Sau}}\ and\ \bibinfo {author} {\bibfnamefont {E.}~\bibnamefont {Demler}},\
  }\href {\doibase 10.1103/PhysRevB.88.205402} {\bibfield  {journal} {\bibinfo
  {journal} {Phys. Rev. B}\ }\textbf {\bibinfo {volume} {88}},\ \bibinfo
  {pages} {205402} (\bibinfo {year} {2013})}\BibitemShut {NoStop}%
\bibitem [{\citenamefont {Pientka}\ \emph {et~al.}(2013)\citenamefont
  {Pientka}, \citenamefont {Glazman},\ and\ \citenamefont {von
  Oppen}}]{Pientka2013}%
  \BibitemOpen
  \bibfield  {author} {\bibinfo {author} {\bibfnamefont {F.}~\bibnamefont
  {Pientka}}, \bibinfo {author} {\bibfnamefont {L.~I.}\ \bibnamefont
  {Glazman}}, \ and\ \bibinfo {author} {\bibfnamefont {F.}~\bibnamefont {von
  Oppen}},\ }\href {\doibase 10.1103/PhysRevB.88.155420} {\bibfield  {journal}
  {\bibinfo  {journal} {Phys. Rev. B}\ }\textbf {\bibinfo {volume} {88}},\
  \bibinfo {pages} {155420} (\bibinfo {year} {2013})}\BibitemShut {NoStop}%
\bibitem [{\citenamefont {Mizushima}\ and\ \citenamefont
  {Sato}(2013)}]{Mizushima2013}%
  \BibitemOpen
  \bibfield  {author} {\bibinfo {author} {\bibfnamefont {T.}~\bibnamefont
  {Mizushima}}\ and\ \bibinfo {author} {\bibfnamefont {M.}~\bibnamefont
  {Sato}},\ }\href {http://stacks.iop.org/1367-2630/15/i=7/a=075010} {\bibfield
   {journal} {\bibinfo  {journal} {New Journal of Physics}\ }\textbf {\bibinfo
  {volume} {15}},\ \bibinfo {pages} {075010} (\bibinfo {year}
  {2013})}\BibitemShut {NoStop}%
\bibitem [{\citenamefont {Wang}\ \emph {et~al.}(2014)\citenamefont {Wang},
  \citenamefont {Huang},\ and\ \citenamefont {Wu}}]{Wang2014}%
  \BibitemOpen
  \bibfield  {author} {\bibinfo {author} {\bibfnamefont {D.}~\bibnamefont
  {Wang}}, \bibinfo {author} {\bibfnamefont {Z.}~\bibnamefont {Huang}}, \ and\
  \bibinfo {author} {\bibfnamefont {C.}~\bibnamefont {Wu}},\ }\href {\doibase
  10.1103/PhysRevB.89.174510} {\bibfield  {journal} {\bibinfo  {journal} {Phys.
  Rev. B}\ }\textbf {\bibinfo {volume} {89}},\ \bibinfo {pages} {174510}
  (\bibinfo {year} {2014})}\BibitemShut {NoStop}%
\bibitem [{\citenamefont {P\"oyh\"onen}\ \emph {et~al.}(2014)\citenamefont
  {P\"oyh\"onen}, \citenamefont {Weststr\"om}, \citenamefont {R\"ontynen},\
  and\ \citenamefont {Ojanen}}]{Poyhonen2014}%
  \BibitemOpen
  \bibfield  {author} {\bibinfo {author} {\bibfnamefont {K.}~\bibnamefont
  {P\"oyh\"onen}}, \bibinfo {author} {\bibfnamefont {A.}~\bibnamefont
  {Weststr\"om}}, \bibinfo {author} {\bibfnamefont {J.}~\bibnamefont
  {R\"ontynen}}, \ and\ \bibinfo {author} {\bibfnamefont {T.}~\bibnamefont
  {Ojanen}},\ }\href {\doibase 10.1103/PhysRevB.89.115109} {\bibfield
  {journal} {\bibinfo  {journal} {Phys. Rev. B}\ }\textbf {\bibinfo {volume}
  {89}},\ \bibinfo {pages} {115109} (\bibinfo {year} {2014})}\BibitemShut
  {NoStop}%
\bibitem [{\citenamefont {Seroussi}\ \emph {et~al.}(2014)\citenamefont
  {Seroussi}, \citenamefont {Berg},\ and\ \citenamefont {Oreg}}]{Seroussi2014}%
  \BibitemOpen
  \bibfield  {author} {\bibinfo {author} {\bibfnamefont {I.}~\bibnamefont
  {Seroussi}}, \bibinfo {author} {\bibfnamefont {E.}~\bibnamefont {Berg}}, \
  and\ \bibinfo {author} {\bibfnamefont {Y.}~\bibnamefont {Oreg}},\ }\href
  {\doibase 10.1103/PhysRevB.89.104523} {\bibfield  {journal} {\bibinfo
  {journal} {Phys. Rev. B}\ }\textbf {\bibinfo {volume} {89}},\ \bibinfo
  {pages} {104523} (\bibinfo {year} {2014})}\BibitemShut {NoStop}%
\bibitem [{\citenamefont {Wakatsuki}\ \emph {et~al.}(2014)\citenamefont
  {Wakatsuki}, \citenamefont {Ezawa},\ and\ \citenamefont
  {Nagaosa}}]{Wakatsuki2014}%
  \BibitemOpen
  \bibfield  {author} {\bibinfo {author} {\bibfnamefont {R.}~\bibnamefont
  {Wakatsuki}}, \bibinfo {author} {\bibfnamefont {M.}~\bibnamefont {Ezawa}}, \
  and\ \bibinfo {author} {\bibfnamefont {N.}~\bibnamefont {Nagaosa}},\ }\href
  {\doibase 10.1103/PhysRevB.89.174514} {\bibfield  {journal} {\bibinfo
  {journal} {Phys. Rev. B}\ }\textbf {\bibinfo {volume} {89}},\ \bibinfo
  {pages} {174514} (\bibinfo {year} {2014})}\BibitemShut {NoStop}%
\bibitem [{\citenamefont {San-Jose}\ \emph {et~al.}(2014)\citenamefont
  {San-Jose}, \citenamefont {Prada},\ and\ \citenamefont
  {Aguado}}]{SanJose2014}%
  \BibitemOpen
  \bibfield  {author} {\bibinfo {author} {\bibfnamefont {P.}~\bibnamefont
  {San-Jose}}, \bibinfo {author} {\bibfnamefont {E.}~\bibnamefont {Prada}}, \
  and\ \bibinfo {author} {\bibfnamefont {R.}~\bibnamefont {Aguado}},\ }\href
  {\doibase 10.1103/PhysRevLett.112.137001} {\bibfield  {journal} {\bibinfo
  {journal} {Phys. Rev. Lett.}\ }\textbf {\bibinfo {volume} {112}},\ \bibinfo
  {pages} {137001} (\bibinfo {year} {2014})}\BibitemShut {NoStop}%
\bibitem [{\citenamefont {Mohanta}\ and\ \citenamefont
  {Taraphder}(2014)}]{Mohanta2014}%
  \BibitemOpen
  \bibfield  {author} {\bibinfo {author} {\bibfnamefont {N.}~\bibnamefont
  {Mohanta}}\ and\ \bibinfo {author} {\bibfnamefont {A.}~\bibnamefont
  {Taraphder}},\ }\href {http://stacks.iop.org/0295-5075/108/i=6/a=60001}
  {\bibfield  {journal} {\bibinfo  {journal} {EPL (Europhysics Letters)}\
  }\textbf {\bibinfo {volume} {108}},\ \bibinfo {pages} {60001} (\bibinfo
  {year} {2014})}\BibitemShut {NoStop}%
\bibitem [{\citenamefont {Ben-Shach}\ \emph {et~al.}(2015)\citenamefont
  {Ben-Shach}, \citenamefont {Haim}, \citenamefont {Appelbaum}, \citenamefont
  {Oreg}, \citenamefont {Yacoby},\ and\ \citenamefont
  {Halperin}}]{BenShach2015}%
  \BibitemOpen
  \bibfield  {author} {\bibinfo {author} {\bibfnamefont {G.}~\bibnamefont
  {Ben-Shach}}, \bibinfo {author} {\bibfnamefont {A.}~\bibnamefont {Haim}},
  \bibinfo {author} {\bibfnamefont {I.}~\bibnamefont {Appelbaum}}, \bibinfo
  {author} {\bibfnamefont {Y.}~\bibnamefont {Oreg}}, \bibinfo {author}
  {\bibfnamefont {A.}~\bibnamefont {Yacoby}}, \ and\ \bibinfo {author}
  {\bibfnamefont {B.~I.}\ \bibnamefont {Halperin}},\ }\href {\doibase
  10.1103/PhysRevB.91.045403} {\bibfield  {journal} {\bibinfo  {journal} {Phys.
  Rev. B}\ }\textbf {\bibinfo {volume} {91}},\ \bibinfo {pages} {045403}
  (\bibinfo {year} {2015})}\BibitemShut {NoStop}%
\bibitem [{\citenamefont {Deng}\ \emph {et~al.}(2012)\citenamefont {Deng},
  \citenamefont {Yu}, \citenamefont {Huang}, \citenamefont {Larsson},
  \citenamefont {Caroff},\ and\ \citenamefont {Xu}}]{Deng2012}%
  \BibitemOpen
  \bibfield  {author} {\bibinfo {author} {\bibfnamefont {M.~T.}\ \bibnamefont
  {Deng}}, \bibinfo {author} {\bibfnamefont {C.~L.}\ \bibnamefont {Yu}},
  \bibinfo {author} {\bibfnamefont {G.~Y.}\ \bibnamefont {Huang}}, \bibinfo
  {author} {\bibfnamefont {M.}~\bibnamefont {Larsson}}, \bibinfo {author}
  {\bibfnamefont {P.}~\bibnamefont {Caroff}}, \ and\ \bibinfo {author}
  {\bibfnamefont {H.~Q.}\ \bibnamefont {Xu}},\ }\href {\doibase
  10.1021/nl303758w} {\bibfield  {journal} {\bibinfo  {journal} {Nano Letters}\
  }\textbf {\bibinfo {volume} {12}},\ \bibinfo {pages} {6414} (\bibinfo {year}
  {2012})},\ \bibinfo {note} {pMID: 23181691}\BibitemShut {NoStop}%
\bibitem [{\citenamefont {Mourik}\ \emph {et~al.}(2012)\citenamefont {Mourik},
  \citenamefont {Zuo}, \citenamefont {Frolov}, \citenamefont {Plissard},
  \citenamefont {Bakkers},\ and\ \citenamefont {Kouwenhoven}}]{Mourik2012}%
  \BibitemOpen
  \bibfield  {author} {\bibinfo {author} {\bibfnamefont {V.}~\bibnamefont
  {Mourik}}, \bibinfo {author} {\bibfnamefont {K.}~\bibnamefont {Zuo}},
  \bibinfo {author} {\bibfnamefont {S.~M.}\ \bibnamefont {Frolov}}, \bibinfo
  {author} {\bibfnamefont {S.~R.}\ \bibnamefont {Plissard}}, \bibinfo {author}
  {\bibfnamefont {E.~P. A.~M.}\ \bibnamefont {Bakkers}}, \ and\ \bibinfo
  {author} {\bibfnamefont {L.~P.}\ \bibnamefont {Kouwenhoven}},\ }\href
  {\doibase 10.1126/science.1222360} {\bibfield  {journal} {\bibinfo  {journal}
  {Science}\ }\textbf {\bibinfo {volume} {336}},\ \bibinfo {pages} {1003}
  (\bibinfo {year} {2012})}\BibitemShut {NoStop}%
\bibitem [{\citenamefont {Das}\ \emph {et~al.}(2012)\citenamefont {Das},
  \citenamefont {Ronen}, \citenamefont {Most}, \citenamefont {Oreg},
  \citenamefont {Heiblum},\ and\ \citenamefont {Shtrikman}}]{Das2012}%
  \BibitemOpen
  \bibfield  {author} {\bibinfo {author} {\bibfnamefont {A.}~\bibnamefont
  {Das}}, \bibinfo {author} {\bibfnamefont {Y.}~\bibnamefont {Ronen}}, \bibinfo
  {author} {\bibfnamefont {Y.}~\bibnamefont {Most}}, \bibinfo {author}
  {\bibfnamefont {Y.}~\bibnamefont {Oreg}}, \bibinfo {author} {\bibfnamefont
  {M.}~\bibnamefont {Heiblum}}, \ and\ \bibinfo {author} {\bibfnamefont
  {H.}~\bibnamefont {Shtrikman}},\ }\href {\doibase 10.1038/nphys2479}
  {\bibfield  {journal} {\bibinfo  {journal} {Nat Phys}\ }\textbf {\bibinfo
  {volume} {8}},\ \bibinfo {pages} {887} (\bibinfo {year} {2012})}\BibitemShut
  {NoStop}%
\bibitem [{\citenamefont {Lee}\ \emph {et~al.}(2014)\citenamefont {Lee},
  \citenamefont {Jiang}, \citenamefont {Houzet}, \citenamefont {Aguado},
  \citenamefont {Lieber},\ and\ \citenamefont {De~Franceschi}}]{Lee2014}%
  \BibitemOpen
  \bibfield  {author} {\bibinfo {author} {\bibfnamefont {E.~J.~H.}\
  \bibnamefont {Lee}}, \bibinfo {author} {\bibfnamefont {X.}~\bibnamefont
  {Jiang}}, \bibinfo {author} {\bibfnamefont {M.}~\bibnamefont {Houzet}},
  \bibinfo {author} {\bibfnamefont {R.}~\bibnamefont {Aguado}}, \bibinfo
  {author} {\bibfnamefont {C.~M.}\ \bibnamefont {Lieber}}, \ and\ \bibinfo
  {author} {\bibfnamefont {S.}~\bibnamefont {De~Franceschi}},\ }\href
  {http://dx.doi.org/10.1038/nnano.2013.267} {\bibfield  {journal} {\bibinfo
  {journal} {Nat Nano}\ }\textbf {\bibinfo {volume} {9}},\ \bibinfo {pages}
  {79} (\bibinfo {year} {2014})}\BibitemShut {NoStop}%
\bibitem [{\citenamefont {Nadj-Perge}\ \emph {et~al.}(2014)\citenamefont
  {Nadj-Perge}, \citenamefont {Drozdov}, \citenamefont {Li}, \citenamefont
  {Chen}, \citenamefont {Jeon}, \citenamefont {Seo}, \citenamefont {MacDonald},
  \citenamefont {Bernevig},\ and\ \citenamefont {Yazdani}}]{NadjPerge2014}%
  \BibitemOpen
  \bibfield  {author} {\bibinfo {author} {\bibfnamefont {S.}~\bibnamefont
  {Nadj-Perge}}, \bibinfo {author} {\bibfnamefont {I.~K.}\ \bibnamefont
  {Drozdov}}, \bibinfo {author} {\bibfnamefont {J.}~\bibnamefont {Li}},
  \bibinfo {author} {\bibfnamefont {H.}~\bibnamefont {Chen}}, \bibinfo {author}
  {\bibfnamefont {S.}~\bibnamefont {Jeon}}, \bibinfo {author} {\bibfnamefont
  {J.}~\bibnamefont {Seo}}, \bibinfo {author} {\bibfnamefont {A.~H.}\
  \bibnamefont {MacDonald}}, \bibinfo {author} {\bibfnamefont {B.~A.}\
  \bibnamefont {Bernevig}}, \ and\ \bibinfo {author} {\bibfnamefont
  {A.}~\bibnamefont {Yazdani}},\ }\href {\doibase 10.1126/science.1259327}
  {\bibfield  {journal} {\bibinfo  {journal} {Science}\ }\textbf {\bibinfo
  {volume} {346}},\ \bibinfo {pages} {602} (\bibinfo {year}
  {2014})}\BibitemShut {NoStop}%
\bibitem [{\citenamefont {Lim}\ \emph {et~al.}(2012)\citenamefont {Lim},
  \citenamefont {Serra}, \citenamefont {L\'opez},\ and\ \citenamefont
  {Aguado}}]{Lim2012}%
  \BibitemOpen
  \bibfield  {author} {\bibinfo {author} {\bibfnamefont {J.~S.}\ \bibnamefont
  {Lim}}, \bibinfo {author} {\bibfnamefont {L.}~\bibnamefont {Serra}}, \bibinfo
  {author} {\bibfnamefont {R.}~\bibnamefont {L\'opez}}, \ and\ \bibinfo
  {author} {\bibfnamefont {R.}~\bibnamefont {Aguado}},\ }\href {\doibase
  10.1103/PhysRevB.86.121103} {\bibfield  {journal} {\bibinfo  {journal} {Phys.
  Rev. B}\ }\textbf {\bibinfo {volume} {86}},\ \bibinfo {pages} {121103}
  (\bibinfo {year} {2012})}\BibitemShut {NoStop}%
\bibitem [{\citenamefont {Lim}\ \emph {et~al.}(2013)\citenamefont {Lim},
  \citenamefont {L\'opez},\ and\ \citenamefont {Serra}}]{Lim2013}%
  \BibitemOpen
  \bibfield  {author} {\bibinfo {author} {\bibfnamefont {J.~S.}\ \bibnamefont
  {Lim}}, \bibinfo {author} {\bibfnamefont {R.}~\bibnamefont {L\'opez}}, \ and\
  \bibinfo {author} {\bibfnamefont {L.}~\bibnamefont {Serra}},\ }\href
  {http://iopscience.iop.org/article/10.1209/0295-5075/103/37004/meta}
  {\bibfield  {journal} {\bibinfo  {journal} {Europhysics Letters}\ }\textbf
  {\bibinfo {volume} {103}},\ \bibinfo {pages} {37004} (\bibinfo {year}
  {2013})}\BibitemShut {NoStop}%
\bibitem [{\citenamefont {Osca}\ and\ \citenamefont {Serra}(2015)}]{Osca2015}%
  \BibitemOpen
  \bibfield  {author} {\bibinfo {author} {\bibfnamefont {J.}~\bibnamefont
  {Osca}}\ and\ \bibinfo {author} {\bibfnamefont {L.}~\bibnamefont {Serra}},\
  }\href {\doibase 10.1103/PhysRevB.91.235417} {\bibfield  {journal} {\bibinfo
  {journal} {Phys. Rev. B}\ }\textbf {\bibinfo {volume} {91}},\ \bibinfo
  {pages} {235417} (\bibinfo {year} {2015})}\BibitemShut {NoStop}%
\bibitem [{\citenamefont {Nijholt}\ and\ \citenamefont
  {Akhmerov}(2016)}]{Nijholt2016}%
  \BibitemOpen
  \bibfield  {author} {\bibinfo {author} {\bibfnamefont {B.}~\bibnamefont
  {Nijholt}}\ and\ \bibinfo {author} {\bibfnamefont {A.~R.}\ \bibnamefont
  {Akhmerov}},\ }\href {\doibase 10.1103/PhysRevB.93.235434} {\bibfield
  {journal} {\bibinfo  {journal} {Phys. Rev. B}\ }\textbf {\bibinfo {volume}
  {93}},\ \bibinfo {pages} {235434} (\bibinfo {year} {2016})}\BibitemShut
  {NoStop}%
\bibitem [{\citenamefont {Kjaergaard}\ \emph {et~al.}(2012)\citenamefont
  {Kjaergaard}, \citenamefont {W\"olms},\ and\ \citenamefont
  {Flensberg}}]{Kjaergaard2012}%
  \BibitemOpen
  \bibfield  {author} {\bibinfo {author} {\bibfnamefont {M.}~\bibnamefont
  {Kjaergaard}}, \bibinfo {author} {\bibfnamefont {K.}~\bibnamefont {W\"olms}},
  \ and\ \bibinfo {author} {\bibfnamefont {K.}~\bibnamefont {Flensberg}},\
  }\href {\doibase 10.1103/PhysRevB.85.020503} {\bibfield  {journal} {\bibinfo
  {journal} {Phys. Rev. B}\ }\textbf {\bibinfo {volume} {85}},\ \bibinfo
  {pages} {020503} (\bibinfo {year} {2012})}\BibitemShut {NoStop}%
\bibitem [{\citenamefont {Loder}\ \emph {et~al.}(2015)\citenamefont {Loder},
  \citenamefont {Kampf},\ and\ \citenamefont {Kopp}}]{Loder2015}%
  \BibitemOpen
  \bibfield  {author} {\bibinfo {author} {\bibfnamefont {F.}~\bibnamefont
  {Loder}}, \bibinfo {author} {\bibfnamefont {A.~P.}\ \bibnamefont {Kampf}}, \
  and\ \bibinfo {author} {\bibfnamefont {T.}~\bibnamefont {Kopp}},\ }\href
  {http://dx.doi.org/10.1038/srep15302} {\bibfield  {journal} {\bibinfo
  {journal} {Scientific Reports}\ }\textbf {\bibinfo {volume} {5}},\ \bibinfo
  {pages} {15302} (\bibinfo {year} {2015})}\BibitemShut {NoStop}%
\bibitem [{\citenamefont {Albrecht}\ \emph {et~al.}(2016)\citenamefont
  {Albrecht}, \citenamefont {Higginbotham}, \citenamefont {Madsen},
  \citenamefont {Kuemmeth}, \citenamefont {Jespersen}, \citenamefont
  {Nyg{\aa}rd}, \citenamefont {Krogstrup},\ and\ \citenamefont
  {Marcus}}]{Albrecht2016}%
  \BibitemOpen
  \bibfield  {author} {\bibinfo {author} {\bibfnamefont {S.~M.}\ \bibnamefont
  {Albrecht}}, \bibinfo {author} {\bibfnamefont {A.~P.}\ \bibnamefont
  {Higginbotham}}, \bibinfo {author} {\bibfnamefont {M.}~\bibnamefont
  {Madsen}}, \bibinfo {author} {\bibfnamefont {F.}~\bibnamefont {Kuemmeth}},
  \bibinfo {author} {\bibfnamefont {T.~S.}\ \bibnamefont {Jespersen}}, \bibinfo
  {author} {\bibfnamefont {J.}~\bibnamefont {Nyg{\aa}rd}}, \bibinfo {author}
  {\bibfnamefont {P.}~\bibnamefont {Krogstrup}}, \ and\ \bibinfo {author}
  {\bibfnamefont {C.~M.}\ \bibnamefont {Marcus}},\ }\href
  {http://dx.doi.org/10.1038/nature17162} {\bibfield  {journal} {\bibinfo
  {journal} {Nature}\ }\textbf {\bibinfo {volume} {531}},\ \bibinfo {pages}
  {206} (\bibinfo {year} {2016})},\ \bibinfo {note} {letter}\BibitemShut
  {NoStop}%
\bibitem [{\citenamefont {Li}\ \emph {et~al.}(2016)\citenamefont {Li},
  \citenamefont {Neupert}, \citenamefont {Bernevig},\ and\ \citenamefont
  {Yazdani}}]{Li2016}%
  \BibitemOpen
  \bibfield  {author} {\bibinfo {author} {\bibfnamefont {J.}~\bibnamefont
  {Li}}, \bibinfo {author} {\bibfnamefont {T.}~\bibnamefont {Neupert}},
  \bibinfo {author} {\bibfnamefont {B.~A.}\ \bibnamefont {Bernevig}}, \ and\
  \bibinfo {author} {\bibfnamefont {A.}~\bibnamefont {Yazdani}},\ }\href
  {\doibase 10.1038/ncomms10395} {\bibfield  {journal} {\bibinfo  {journal}
  {Nature Communications}\ }\textbf {\bibinfo {volume} {7}},\ \bibinfo {pages}
  {10395 EP } (\bibinfo {year} {2016})}\BibitemShut {NoStop}%
\bibitem [{\citenamefont {Sticlet}\ \emph {et~al.}(2012)\citenamefont
  {Sticlet}, \citenamefont {Bena},\ and\ \citenamefont {Simon}}]{Sticlet2012}%
  \BibitemOpen
  \bibfield  {author} {\bibinfo {author} {\bibfnamefont {D.}~\bibnamefont
  {Sticlet}}, \bibinfo {author} {\bibfnamefont {C.}~\bibnamefont {Bena}}, \
  and\ \bibinfo {author} {\bibfnamefont {P.}~\bibnamefont {Simon}},\ }\href
  {\doibase 10.1103/PhysRevLett.108.096802} {\bibfield  {journal} {\bibinfo
  {journal} {Phys. Rev. Lett.}\ }\textbf {\bibinfo {volume} {108}},\ \bibinfo
  {pages} {096802} (\bibinfo {year} {2012})}\BibitemShut {NoStop}%
\bibitem [{\citenamefont {Sedlmayr}\ \emph {et~al.}(2015)\citenamefont
  {Sedlmayr}, \citenamefont {Aguiar-Hualde},\ and\ \citenamefont
  {Bena}}]{Sedlmayr2015a}%
  \BibitemOpen
  \bibfield  {author} {\bibinfo {author} {\bibfnamefont {N.}~\bibnamefont
  {Sedlmayr}}, \bibinfo {author} {\bibfnamefont {J.~M.}\ \bibnamefont
  {Aguiar-Hualde}}, \ and\ \bibinfo {author} {\bibfnamefont {C.}~\bibnamefont
  {Bena}},\ }\href {\doibase 10.1103/PhysRevB.91.115415} {\bibfield  {journal}
  {\bibinfo  {journal} {Phys. Rev. B}\ }\textbf {\bibinfo {volume} {91}},\
  \bibinfo {pages} {115415} (\bibinfo {year} {2015})}\BibitemShut {NoStop}%
\bibitem [{\citenamefont {Sedlmayr}\ and\ \citenamefont
  {Bena}(2015)}]{Sedlmayr2015b}%
  \BibitemOpen
  \bibfield  {author} {\bibinfo {author} {\bibfnamefont {N.}~\bibnamefont
  {Sedlmayr}}\ and\ \bibinfo {author} {\bibfnamefont {C.}~\bibnamefont
  {Bena}},\ }\href {\doibase 10.1103/PhysRevB.92.115115} {\bibfield  {journal}
  {\bibinfo  {journal} {Phys. Rev. B}\ }\textbf {\bibinfo {volume} {92}},\
  \bibinfo {pages} {115115} (\bibinfo {year} {2015})}\BibitemShut {NoStop}%
\bibitem [{\citenamefont {Sedlmayr}\ \emph {et~al.}(2016)\citenamefont
  {Sedlmayr}, \citenamefont {Aguiar-Hualde},\ and\ \citenamefont
  {Bena}}]{Sedlmayr2016}%
  \BibitemOpen
  \bibfield  {author} {\bibinfo {author} {\bibfnamefont {N.}~\bibnamefont
  {Sedlmayr}}, \bibinfo {author} {\bibfnamefont {J.~M.}\ \bibnamefont
  {Aguiar-Hualde}}, \ and\ \bibinfo {author} {\bibfnamefont {C.}~\bibnamefont
  {Bena}},\ }\href {\doibase 10.1103/PhysRevB.93.155425} {\bibfield  {journal}
  {\bibinfo  {journal} {Phys. Rev. B}\ }\textbf {\bibinfo {volume} {93}},\
  \bibinfo {pages} {155425} (\bibinfo {year} {2016})}\BibitemShut {NoStop}%
\bibitem [{\citenamefont {Altland}\ and\ \citenamefont
  {Zirnbauer}(1997)}]{Altland1997}%
  \BibitemOpen
  \bibfield  {author} {\bibinfo {author} {\bibfnamefont {A.}~\bibnamefont
  {Altland}}\ and\ \bibinfo {author} {\bibfnamefont {M.~R.}\ \bibnamefont
  {Zirnbauer}},\ }\href {\doibase 10.1103/PhysRevB.55.1142} {\bibfield
  {journal} {\bibinfo  {journal} {Phys. Rev. B}\ }\textbf {\bibinfo {volume}
  {55}},\ \bibinfo {pages} {1142} (\bibinfo {year} {1997})}\BibitemShut
  {NoStop}%
\bibitem [{\citenamefont {Sato}\ and\ \citenamefont
  {Fujimoto}(2010)}]{Sato2010}%
  \BibitemOpen
  \bibfield  {author} {\bibinfo {author} {\bibfnamefont {M.}~\bibnamefont
  {Sato}}\ and\ \bibinfo {author} {\bibfnamefont {S.}~\bibnamefont
  {Fujimoto}},\ }\href {\doibase 10.1103/PhysRevLett.105.217001} {\bibfield
  {journal} {\bibinfo  {journal} {Phys. Rev. Lett.}\ }\textbf {\bibinfo
  {volume} {105}},\ \bibinfo {pages} {217001} (\bibinfo {year}
  {2010})}\BibitemShut {NoStop}%
\bibitem [{\citenamefont {Qi}\ \emph {et~al.}(2010)\citenamefont {Qi},
  \citenamefont {Hughes},\ and\ \citenamefont {Zhang}}]{Qi2010}%
  \BibitemOpen
  \bibfield  {author} {\bibinfo {author} {\bibfnamefont {X.-L.}\ \bibnamefont
  {Qi}}, \bibinfo {author} {\bibfnamefont {T.~L.}\ \bibnamefont {Hughes}}, \
  and\ \bibinfo {author} {\bibfnamefont {S.-C.}\ \bibnamefont {Zhang}},\ }\href
  {\doibase 10.1103/PhysRevB.82.184516} {\bibfield  {journal} {\bibinfo
  {journal} {Phys. Rev. B}\ }\textbf {\bibinfo {volume} {82}},\ \bibinfo
  {pages} {184516} (\bibinfo {year} {2010})}\BibitemShut {NoStop}%
\bibitem [{\citenamefont {Wu}\ \emph {et~al.}(2012)\citenamefont {Wu},
  \citenamefont {Liang}, \citenamefont {Wang},\ and\ \citenamefont
  {Hu}}]{Wu2012}%
  \BibitemOpen
  \bibfield  {author} {\bibinfo {author} {\bibfnamefont {L.-H.}\ \bibnamefont
  {Wu}}, \bibinfo {author} {\bibfnamefont {Q.-F.}\ \bibnamefont {Liang}},
  \bibinfo {author} {\bibfnamefont {Z.}~\bibnamefont {Wang}}, \ and\ \bibinfo
  {author} {\bibfnamefont {X.}~\bibnamefont {Hu}},\ }\href
  {http://stacks.iop.org/1742-6596/393/i=1/a=012018} {\bibfield  {journal}
  {\bibinfo  {journal} {Journal of Physics: Conference Series}\ }\textbf
  {\bibinfo {volume} {393}},\ \bibinfo {pages} {012018} (\bibinfo {year}
  {2012})}\BibitemShut {NoStop}%
\bibitem [{\citenamefont {Daido}\ and\ \citenamefont
  {Yanase}(2017)}]{Daido2017}%
  \BibitemOpen
  \bibfield  {author} {\bibinfo {author} {\bibfnamefont {A.}~\bibnamefont
  {Daido}}\ and\ \bibinfo {author} {\bibfnamefont {Y.}~\bibnamefont {Yanase}},\
  }\href {\doibase 10.1103/PhysRevB.95.134507} {\bibfield  {journal} {\bibinfo
  {journal} {Phys. Rev. B}\ }\textbf {\bibinfo {volume} {95}},\ \bibinfo
  {pages} {134507} (\bibinfo {year} {2017})}\BibitemShut {NoStop}%
\bibitem [{\citenamefont {He}\ \emph {et~al.}(2017)\citenamefont {He},
  \citenamefont {Pan}, \citenamefont {Stern}, \citenamefont {Burks},
  \citenamefont {Che}, \citenamefont {Yin}, \citenamefont {Wang}, \citenamefont
  {Lian}, \citenamefont {Zhou}, \citenamefont {Choi}, \citenamefont {Murata},
  \citenamefont {Kou}, \citenamefont {Chen}, \citenamefont {Nie}, \citenamefont
  {Shao}, \citenamefont {Fan}, \citenamefont {Zhang}, \citenamefont {Liu},
  \citenamefont {Xia},\ and\ \citenamefont {Wang}}]{He2017}%
  \BibitemOpen
  \bibfield  {author} {\bibinfo {author} {\bibfnamefont {Q.~L.}\ \bibnamefont
  {He}}, \bibinfo {author} {\bibfnamefont {L.}~\bibnamefont {Pan}}, \bibinfo
  {author} {\bibfnamefont {A.~L.}\ \bibnamefont {Stern}}, \bibinfo {author}
  {\bibfnamefont {E.~C.}\ \bibnamefont {Burks}}, \bibinfo {author}
  {\bibfnamefont {X.}~\bibnamefont {Che}}, \bibinfo {author} {\bibfnamefont
  {G.}~\bibnamefont {Yin}}, \bibinfo {author} {\bibfnamefont {J.}~\bibnamefont
  {Wang}}, \bibinfo {author} {\bibfnamefont {B.}~\bibnamefont {Lian}}, \bibinfo
  {author} {\bibfnamefont {Q.}~\bibnamefont {Zhou}}, \bibinfo {author}
  {\bibfnamefont {E.~S.}\ \bibnamefont {Choi}}, \bibinfo {author}
  {\bibfnamefont {K.}~\bibnamefont {Murata}}, \bibinfo {author} {\bibfnamefont
  {X.}~\bibnamefont {Kou}}, \bibinfo {author} {\bibfnamefont {Z.}~\bibnamefont
  {Chen}}, \bibinfo {author} {\bibfnamefont {T.}~\bibnamefont {Nie}}, \bibinfo
  {author} {\bibfnamefont {Q.}~\bibnamefont {Shao}}, \bibinfo {author}
  {\bibfnamefont {Y.}~\bibnamefont {Fan}}, \bibinfo {author} {\bibfnamefont
  {S.-C.}\ \bibnamefont {Zhang}}, \bibinfo {author} {\bibfnamefont
  {K.}~\bibnamefont {Liu}}, \bibinfo {author} {\bibfnamefont {J.}~\bibnamefont
  {Xia}}, \ and\ \bibinfo {author} {\bibfnamefont {K.~L.}\ \bibnamefont
  {Wang}},\ }\href {\doibase 10.1126/science.aag2792} {\bibfield  {journal}
  {\bibinfo  {journal} {Science}\ }\textbf {\bibinfo {volume} {357}},\ \bibinfo
  {pages} {294} (\bibinfo {year} {2017})}\BibitemShut {NoStop}%
\bibitem [{mat()}]{matq}%
  \BibitemOpen
  \href@noop {} {\emph {\bibinfo {title} {MatQ}}},\ \bibinfo {address}
  {www.icmm.csic.es/sanjose/MathQ/MathQ.html}\BibitemShut {NoStop}%
\bibitem [{\citenamefont {Teo}\ and\ \citenamefont {Kane}(2010)}]{Teo2010}%
  \BibitemOpen
  \bibfield  {author} {\bibinfo {author} {\bibfnamefont {J.~C.~Y.}\
  \bibnamefont {Teo}}\ and\ \bibinfo {author} {\bibfnamefont {C.~L.}\
  \bibnamefont {Kane}},\ }\href {\doibase 10.1103/PhysRevB.82.115120}
  {\bibfield  {journal} {\bibinfo  {journal} {Phys. Rev. B}\ }\textbf {\bibinfo
  {volume} {82}},\ \bibinfo {pages} {115120} (\bibinfo {year}
  {2010})}\BibitemShut {NoStop}%
\bibitem [{\citenamefont {Matsuura}\ \emph {et~al.}(2013)\citenamefont
  {Matsuura}, \citenamefont {Chang}, \citenamefont {Schnyder},\ and\
  \citenamefont {Ryu}}]{Matsuura2013}%
  \BibitemOpen
  \bibfield  {author} {\bibinfo {author} {\bibfnamefont {S.}~\bibnamefont
  {Matsuura}}, \bibinfo {author} {\bibfnamefont {P.-Y.}\ \bibnamefont {Chang}},
  \bibinfo {author} {\bibfnamefont {A.~P.}\ \bibnamefont {Schnyder}}, \ and\
  \bibinfo {author} {\bibfnamefont {S.}~\bibnamefont {Ryu}},\ }\href
  {http://stacks.iop.org/1367-2630/15/i=6/a=065001} {\bibfield  {journal}
  {\bibinfo  {journal} {New Journal of Physics}\ }\textbf {\bibinfo {volume}
  {15}},\ \bibinfo {pages} {065001} (\bibinfo {year} {2013})}\BibitemShut
  {NoStop}%
\bibitem [{\citenamefont {Deng}\ \emph {et~al.}(2014)\citenamefont {Deng},
  \citenamefont {Ortiz}, \citenamefont {Poudel},\ and\ \citenamefont
  {Viola}}]{Deng2014}%
  \BibitemOpen
  \bibfield  {author} {\bibinfo {author} {\bibfnamefont {S.}~\bibnamefont
  {Deng}}, \bibinfo {author} {\bibfnamefont {G.}~\bibnamefont {Ortiz}},
  \bibinfo {author} {\bibfnamefont {A.}~\bibnamefont {Poudel}}, \ and\ \bibinfo
  {author} {\bibfnamefont {L.}~\bibnamefont {Viola}},\ }\href {\doibase
  10.1103/PhysRevB.89.140507} {\bibfield  {journal} {\bibinfo  {journal} {Phys.
  Rev. B}\ }\textbf {\bibinfo {volume} {89}},\ \bibinfo {pages} {140507}
  (\bibinfo {year} {2014})}\BibitemShut {NoStop}%
\bibitem [{\citenamefont {Baum}\ \emph
  {et~al.}(2015{\natexlab{a}})\citenamefont {Baum}, \citenamefont {Posske},
  \citenamefont {Fulga}, \citenamefont {Trauzettel},\ and\ \citenamefont
  {Stern}}]{Baum2015a}%
  \BibitemOpen
  \bibfield  {author} {\bibinfo {author} {\bibfnamefont {Y.}~\bibnamefont
  {Baum}}, \bibinfo {author} {\bibfnamefont {T.}~\bibnamefont {Posske}},
  \bibinfo {author} {\bibfnamefont {I.~C.}\ \bibnamefont {Fulga}}, \bibinfo
  {author} {\bibfnamefont {B.}~\bibnamefont {Trauzettel}}, \ and\ \bibinfo
  {author} {\bibfnamefont {A.}~\bibnamefont {Stern}},\ }\href {\doibase
  10.1103/PhysRevLett.114.136801} {\bibfield  {journal} {\bibinfo  {journal}
  {Phys. Rev. Lett.}\ }\textbf {\bibinfo {volume} {114}},\ \bibinfo {pages}
  {136801} (\bibinfo {year} {2015}{\natexlab{a}})}\BibitemShut {NoStop}%
\bibitem [{\citenamefont {Baum}\ \emph
  {et~al.}(2015{\natexlab{b}})\citenamefont {Baum}, \citenamefont {Posske},
  \citenamefont {Fulga}, \citenamefont {Trauzettel},\ and\ \citenamefont
  {Stern}}]{Baum2015b}%
  \BibitemOpen
  \bibfield  {author} {\bibinfo {author} {\bibfnamefont {Y.}~\bibnamefont
  {Baum}}, \bibinfo {author} {\bibfnamefont {T.}~\bibnamefont {Posske}},
  \bibinfo {author} {\bibfnamefont {I.~C.}\ \bibnamefont {Fulga}}, \bibinfo
  {author} {\bibfnamefont {B.}~\bibnamefont {Trauzettel}}, \ and\ \bibinfo
  {author} {\bibfnamefont {A.}~\bibnamefont {Stern}},\ }\href {\doibase
  10.1103/PhysRevB.92.045128} {\bibfield  {journal} {\bibinfo  {journal} {Phys.
  Rev. B}\ }\textbf {\bibinfo {volume} {92}},\ \bibinfo {pages} {045128}
  (\bibinfo {year} {2015}{\natexlab{b}})}\BibitemShut {NoStop}%
\bibitem [{\citenamefont {Gibertini}\ \emph {et~al.}(2012)\citenamefont
  {Gibertini}, \citenamefont {Taddei}, \citenamefont {Polini},\ and\
  \citenamefont {Fazio}}]{Gibertini2012}%
  \BibitemOpen
  \bibfield  {author} {\bibinfo {author} {\bibfnamefont {M.}~\bibnamefont
  {Gibertini}}, \bibinfo {author} {\bibfnamefont {F.}~\bibnamefont {Taddei}},
  \bibinfo {author} {\bibfnamefont {M.}~\bibnamefont {Polini}}, \ and\ \bibinfo
  {author} {\bibfnamefont {R.}~\bibnamefont {Fazio}},\ }\href {\doibase
  10.1103/PhysRevB.85.144525} {\bibfield  {journal} {\bibinfo  {journal} {Phys.
  Rev. B}\ }\textbf {\bibinfo {volume} {85}},\ \bibinfo {pages} {144525}
  (\bibinfo {year} {2012})}\BibitemShut {NoStop}%
\end{thebibliography}%

\widetext
\appendix

\section{Topological invariant calculation}

We write the Bogoliubov-de Gennes Hamiltonian on a square lattice in the Nambu basis $\left\{ \psi_\uparrow,\psi_\downarrow, \psi^\dag_\downarrow, -\psi^\dag_\uparrow \right\}$ as:
\begin{align}
\mathcal{H}(k) = 
	\bpm 
		f(k) & \mathcal{L}(k) & \Delta & 0  \\ 
		\mathcal{L}^*(k) & f(k) & 0 & \Delta  \\
		\Delta & 0 & -f(k) & \mathcal{L}^*(-k) \\
		0 & \Delta & \mathcal{L}(-k) & -f(k)
	\epm +
	\bpm
		B_z & B_x - i B_y & 0 & 0 \\
		B_x + i B_y & -B_z & 0 & 0 \\
		0 & 0 & B_z & B_x - i B_y \\
		0 & 0 & B_x + i B_y & -B_z 
	\epm
\label{H1D}
\end{align}
where $\bs B = (B_x, B_y, B_z)$ is the magnetic field, $f(k)$ is the dimension-dependent dispersion for electrons on the lattice with a chemical potential $\mu$ and a hopping parameter $t$, while $\mathcal{L}(k) $ is the Rashba spin-orbit coupling term, which also depends on the dimensionality of the lattice. We disregard the orbital effects of the magnetic field. 
The Hamiltonian given by Eq.~(\ref{H1D}) is particle-hole symmetric, i.e.
\begin{align}
\Xi \mathcal{H}(k) \Xi^{-1} = -\mathcal{H}(-k), \quad \Xi \equiv \sigma_y \otimes \tau_y \,\mathcal{K},
\end{align}
where $\mathcal{K}$ is the complex conjugate operator, and we set $\Lambda \equiv \sigma_y \otimes \tau_y$.\\ 

\subsection{1D wires}

We consider a 1D superconducting nanowire with Rashba spin-orbit coupling and with an arbitrary direction of the magnetic field. We thus have $f(k) \equiv -2t\cos k-\mu$ and $\mathcal{L}(k) \equiv i\lambda \sin k $. As long as the magnetic field is not collinear with the spin-orbit coupling ($B_z$ or $B_x$, but not $B_y$), the system stays gapful and thus topological invariants are well-defined. Time-reversal symmetry is broken in the presence of a magnetic field, therefore we expect to have a $\mathbb{Z}_2$ topological invariant in accordance with the classification in Ref.~[\onlinecite{Altland1997}]. On the contrary, if we consider the case of a non-zero $B_y$ along the y-axis, we end up having a gapless trivial phase. Thus, below we set $B_y=0$.

\begin{figure}[h!]
	\includegraphics*[width = 0.3\columnwidth]{1Dwire-TI.pdf}
	\includegraphics*[width = 0.3\columnwidth]{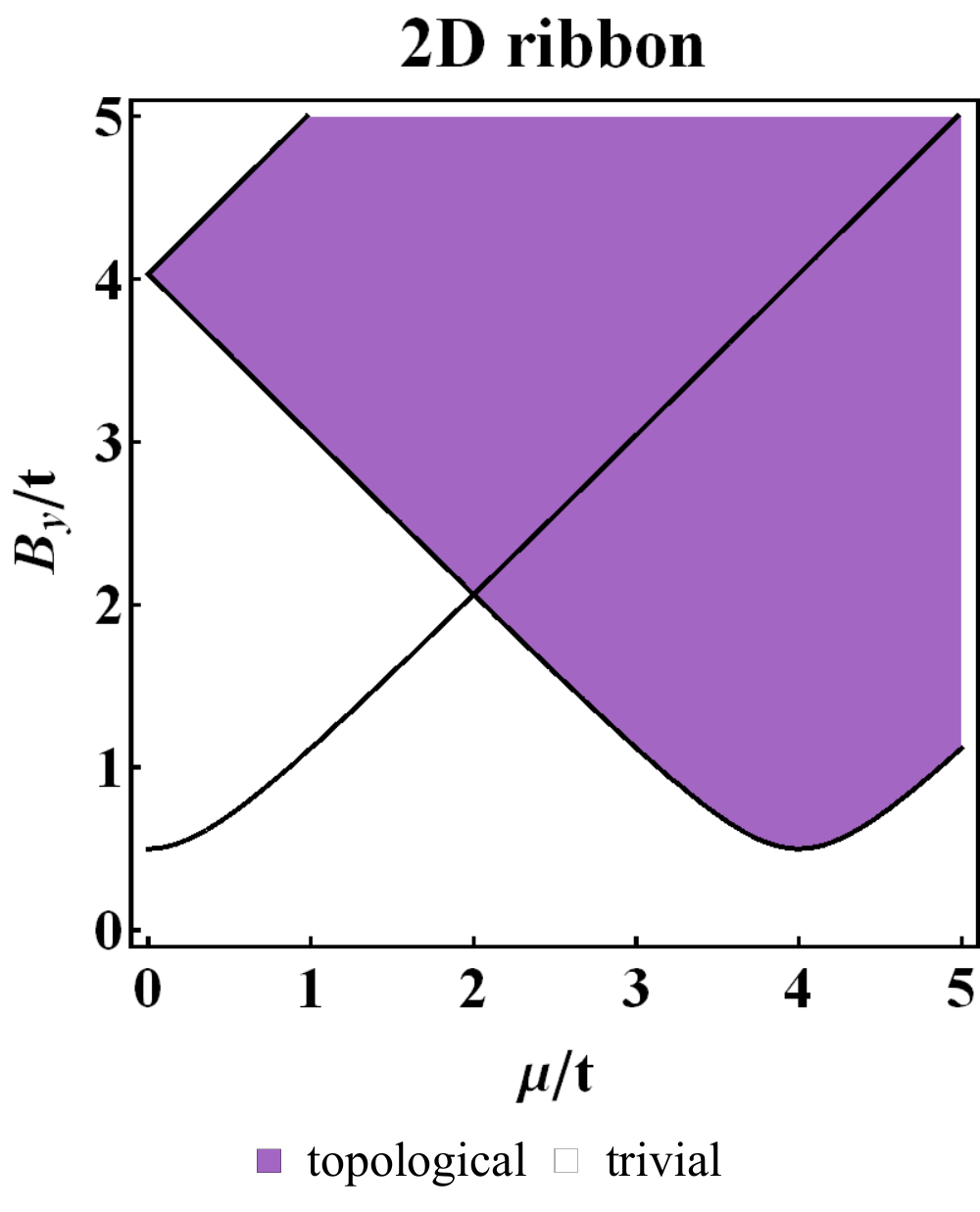}
	\caption{(color online) The phase diagram of a 1D nanowire (left panel) and of an infinite ribbon (right panel) as a function of the chemical potential $\mu$ and the magnetic field $B$, obtained using a topological invariant calculation. In the case of a wire the magnetic field is either along the wire $B=B_x$ or perpendicular to the wire $B=B_z$, whereas in the case of an infinite ribbon it is perpendicular to the edge of the ribbon, i.e. $B=B_y$. The black lines correspond to the values for which the topological invariant $\delta=0$. The topological regions are colored in violet. We set $\Delta=0.2t, \lambda_x=0.5t, \lambda_y=0t$ for the left panel and $\Delta=0.2t, \lambda_x=\lambda_y=0.5t$ for the right one.}
	\label{phasediagram1D2D}
\end{figure}

To compute the topological invariant we seek the Gamma-points of the Hamiltonian (\ref{H1D}), i.e. the points for which $\mathcal{L}(k) = 0$. It is easy to see that within the first Brillouin zone we have two such points:
$
\Gamma_1 = 0, \; \Gamma_2 = \pi.
$
At each of these points one can define a skew-symmetric matrix $W(k = \Gamma_i) = \mathcal{H}(k = \Gamma_i)\Lambda$ with an associated Pfaffian. The topological invariant is given by
\begin{align}
\delta = \prod\limits_{k = \bs\Gamma_i} \frac{\sqrt{\det\left[W(k)\right]}}{\Pf\left[W(k)\right]}.
\end{align}
This expression can be simplified using the identity $(\Pf A)^2 = \det A$, thus yielding:
\begin{align}
\delta = \prod\limits_{k = \Gamma_i} \sgn \Pf\left[W(k)\right] = \sgn \Pf\left[W(0)\right] \sgn \Pf\left[W(\pi)\right].
\end{align}
We compute the corresponding Pfaffians
\begin{align*}
\Pf\left[W(0)\right] = B^2 - \Delta^2 - (\mu +2t)^2, \quad 
\Pf\left[W(\pi)\right] = B^2 - \Delta^2 - (\mu -2t)^2,
\end{align*}
where $B^2 \equiv |\bs B|^2 = B_x^2 + B_z^2$. Therefore, the topological invariant is given by:
\begin{align}
\delta = \sgn \Big[(B^2 - \Delta^2 - (\mu+2t)^2)(B^2 - \Delta^2 - (\mu -2t)^2) \Big].
\end{align}
The corresponding phase diagram is given in Fig. \ref{1wire-PhD} as well as on the left panel of Fig.~\ref{phasediagram1D2D}.

\subsection{2D systems}
For this system $f(\bs k) \equiv -2t(\cos k_x + \cos k_y)-\mu$ and $\mathcal{L}(\bs k) \equiv i\lambda (\sin k_x - i \sin k_y)$ is the Rashba spin-orbit coupling term. Despite the fact that in parallel magnetic fields the system can become gapless for certain regions in the parameter space, below we compute {\it formally} a $\mathbb{Z}_2$ topological invariant. This invariant should indicate the parity of the number of Majorana modes arising at the edges introduced into the system. To compute the topological invariant we find the Gamma-points of the Hamiltonian (\ref{H1D}), i.e. the points where $\mathcal{L}(\bs k) = 0$. Within the first Brillouin zone there are  four such points:
$
\bs\Gamma_1 = (0,0), \; \bs\Gamma_2 = (\pi,0),\; \bs\Gamma_3 = (0,\pi),\; \bs\Gamma_4 = (\pi,\pi).
$
At each of these points one can define a skew-symmetric matrix $W(\bs k = \bs\Gamma_i) = \mathcal{H}(\bs k = \Gamma_i)\Lambda$ with an associated Pfaffian. Exactly as in the previous subsection the topological invariant is given by
\begin{align}
\delta = \prod\limits_{\bs k = \bs\Gamma_i} \sgn \Pf\left[W(\bs k)\right].
\end{align}
Computing the corresponding Pfaffians we get:
\begin{align*}
\Pf\left[W(\bs\Gamma_1)\right] = B^2 - \Delta_s^2 - (\mu + 4t)^2,\;
\Pf\left[W(\bs\Gamma_2)\right] = \Pf\left[W(\bs\Gamma_3)\right] = B^2 - \Delta^2 - \mu^2,\;
\Pf\left[W(\bs\Gamma_4)\right] = B^2 - \Delta^2 - (\mu - 4t)^2,
\end{align*}
where $B^2 \equiv |\bs B|^2 = B_x^2 + B_y^2 + B_z^2$. Therefore, the topological invariant is given by:
\begin{align}
\delta = (B^2 - \Delta^2 - \mu^2)^2 \sgn \Big[ (B^2 - \Delta^2 - (\mu + 4t)^2)(B^2 - \Delta^2 - (\mu - 4t)^2) \Big].
\end{align}
The corresponding phase diagram is given in the right panel of the Fig.~\ref{phasediagram1D2D}. It is clear that there is no difference between the cases of a magnetic field perpendicular to the plane and an in-plane magnetic field perpendicular to the edges of the ribbon (note however that an in-plane magnetic field parallel to the edges of the ribbon does not give rise to a topological phase for any region in the parameter space). Despite the closing of the gap, this calculation yields the correct result for the parity of the number of Majorana modes (compare with the results of the numerical simulations presented in Fig.~\ref{2Dribbon}). 

\subsection{Finite-size strips}

We follow the section about finite-size strips in Ref.~[\onlinecite{Sedlmayr2016}], but instead of considering a magnetic field perpendicular to the plane of the  system, we consider also non-zero in-plane components of the magnetic field, $B_x$ and $B_y$. The Fourier-transformed Hamiltonian of this system in the case of $N_y$ coupled wires can be written in the Nambu basis $\left\{ \psi_\uparrow,\psi_\downarrow, \psi^\dag_\downarrow, -\psi^\dag_\uparrow \right\}$ in the following manner:
\begin{eqnarray}
\mathcal{H}( k) &=&
	\bpm 
		f( k)\mathbb{I}_{N_y} - t M_{1N_y} & \mathcal{L}( k)\mathbb{I}_{N_y} + i \lambda M_{2N_y} & \Delta \mathbb{I}_{N_y} & \hat{0}  \\ 
		\left[ -\mathcal{L}( k)\mathbb{I}_{N_y} - i \lambda M_{2N_y} \right]^{\T} & f( k)\mathbb{I}_{N_y} - t M_{1N_y} & \hat{0} & \Delta \mathbb{I}_{N_y} \\
		\Delta \mathbb{I}_{N_y} & \hat{0} & -f( k)\mathbb{I}_{N_y} + t M_{1N_y} & \left[\mathcal{L}(- k)\mathbb{I}_{N_y} + i \lambda M_{2N_y} \right]^{\T} \\
		\hat{0} & \Delta \mathbb{I}_{N_y} & -\mathcal{L}(- k)\mathbb{I}_{N_y} - i \lambda M_{2N_y} & -f( k)\mathbb{I}_{N_y} + t M_{1N_y}
	\epm + \phantom{aaaa}\\ 
&+&	\bpm
		B_z \mathbb{I}_{N_y} & (B_x - i B_y)\mathbb{I}_{N_y} & \hat{0} & \hat{0} \\
		(B_x + i B_y) \mathbb{I}_{N_y} & -B_z \mathbb{I}_{N_y} & \hat{0} & \hat{0} \\
		\hat{0} & \hat{0} & B_z \mathbb{I}_{N_y} & (B_x - i B_y) \mathbb{I}_{N_y}\\
		\hat{0} & \hat{0} & (B_x + i B_y) \mathbb{I}_{N_y} & -B_z \mathbb{I}_{N_y}
	\epm, 
\label{H2Dq1D}
\end{eqnarray}
where the $N_y \times N_y$ square matrices $\mathbb{I}_{N_y}, M_{1N_y}, M_{2N_y}$ act in the sublattice space and are defined as follows:
$$
\left[\mathbb{I}_{N_y}\right]_{ij} = \delta_{ij}, \quad \left[M_{1N_y}\right]_{ij} =  \delta_{i,j-1} + \delta_{i-1,j}, \quad \left[M_{2N_y}\right]_{ij} =  -\delta_{i,j-1} + \delta_{i-1,j}, \quad \forall i,j \in \overline{1,N_y},
$$
with $\delta$ denoting the Kronecker delta. We define also the spectrum of the free electrons $ f(k) = -2t\cos k - \mu $, and the spin-orbit coupling term $\mathcal{L}(k) = -2i \lambda\sin k $.
In this system the PHS operator can be written as follows:
\begin{align}
\Xi \equiv \Lambda \mathcal{K} \equiv 
	\begin{pmatrix}
		\hat{0} & \hat{0} & \hat{0} & -\mathbb{I}_{N_y} \\
		\hat{0} & \hat{0} & \mathbb{I}_{N_y} & \hat{0} \\
		\hat{0} & \mathbb{I}_{N_y} & \hat{0} & \hat{0} \\
		-\mathbb{I}_{N_y} & \hat{0} & \hat{0} & \hat{0} 
	\end{pmatrix} \; \mathcal{K},
\end{align}
where $\mathcal{K}$ is the complex conjugation operator. We take the same path as in the previous subsections: firstly, we find the Gamma-points of the Hamiltonian (\ref{H2Dq1D}), i.e. the points where $\mathcal{L}(k)=0$. We find only two such points in the first Brillouin zone:
$
\Gamma_1 = 0,\; \Gamma_2 = \pi.
$
A skew-symmetric matrix $W(k=\Gamma_i) \equiv \mathcal{H}(k=\Gamma_i)\Lambda$ is defined at each of those points along with an associated Pfaffian. Exactly as before, the topological invariant is given by
\begin{align}
\delta = \sgn \Pf \left[W(0)\right] \sgn \Pf \left[W(\pi)\right]
\label{TI2Dq1Dformula}
\end{align}
We do not give here the analytical expressions for $\delta$ since they are quite cumbersome. To obtain the topological phase diagram we plot in the right column of Fig. \ref{disorderq1D} the topological invariant as given by Eq.~(\ref{TI2Dq1Dformula}), as a function of an in-plane magnetic field $B_x$ and the chemical potential $\mu$.

\end{document}